%% file: main.tex
\newcommand\arxivversion

\ifdefined\JACM
\documentclass[format=acmsmall, review=false, screen=true]{acmart}
\acmJournal{JACM}

\received{February 2018}

\else

\RequirePackage{fix-cm}
\documentclass[10pt]{article}
\usepackage{fullpage}

\fi

\usepackage[latin1]{inputenc}
\usepackage[english]{babel}

\usepackage{graphicx}
\DeclareGraphicsExtensions{.jpg, .pdf, .ps}
\usepackage{xcolor}
\usepackage{subfig}

\usepackage{array}

\usepackage{amsthm}
\usepackage{amssymb}
\usepackage{amsmath}
\usepackage{enumitem}

\usepackage{multirow}
\usepackage{afterpage}
\usepackage[normalem]{ulem}
\usepackage{makecell}
\usepackage{wasysym}
\usepackage{tikz} 
\usetikzlibrary{shapes.geometric}
\usetikzlibrary{arrows,positioning,calc,fit}

\usepackage{url}

\ifdefined\JACM
\else
\usepackage[pdftex,bookmarks=true,colorlinks=true,linkcolor=blue,citecolor=blue,urlcolor=blue]{hyperref} 
\fi

\newcommand{\delete}[1]{}
\newcommand{\techreport}[1]{}
\newcommand{\semterm}[1]{\langle\!\langle #1 \rangle\!\rangle}

\input{line_count}

\input{macros}

\newtheorem{theorem}{Theorem}
\newtheorem{lemma}{Lemma}
\newtheorem{corollary}{Corollary}
\newtheorem{definition}{Definition}

\ifx\proof\undefined
\newenvironment{proof}[1][\protect\proofname]{\par
\normalfont\topsep6\p@\@plus6\p@\relax
\trivlist
\itemindent\parindent
\item[\hskip\labelsep\scshape #1]\ignorespaces
}{%
\endtrivlist\@endpefalse
}
\providecommand{\proofname}{Proof}
\fi

\newcommand\Item[1][]{%
  \ifx\relax#1\relax  \item \else \item[#1] \fi
  \abovedisplayskip=0pt\abovedisplayshortskip=0pt~\vspace*{-\baselineskip}}

\ifdefined\JACM

\begin{document}

\title{The Refinement Calculus of Reactive Systems}
\titlenote{
The bulk of this work was done while Preoteasa and Dragomir were at Aalto
University. Preoteasa is currently at Space Systems Finland.
Dragomir is currently at Verimag. Tripakis is affiliated with
Aalto University and the University of California at Berkeley.
This work was partially supported by the Academy of Finland and the U.S. National Science Foundation (awards \#1329759 and \#1139138).
Dragomir was supported during the writing of this paper by the Horizon 2020 Programme Strategic Research Cluster (grant agreement \#730080 and \#730086).}

\author{Viorel Preoteasa}
\author{Iulia Dragomir}
\author{Stavros Tripakis}

\input{abstract}




\maketitle

\renewcommand{\shortauthors}{V. Preoteasa et al.}

\else
\title{The Refinement Calculus of Reactive Systems\thanks{
The bulk of this work was done while Preoteasa and Dragomir were at Aalto
University. Preoteasa is currently at Space Systems Finland.
Dragomir is currently at Verimag. Tripakis is affiliated with
Aalto University and the University of California at Berkeley.
This work was partially supported by the Academy of Finland and the U.S. National Science Foundation (awards \#1329759 and \#1139138).
Dragomir was supported during the writing of this paper by the Horizon 2020 Programme Strategic Research Cluster (grant agreement \#730080 and \#730086).}}

\author{Viorel Preoteasa \and Iulia Dragomir \and Stavros Tripakis}




\begin{document}

\maketitle

\input{abstract}
\fi

\ifdefined\JACM
\setcounter{tocdepth}{2}
\else
\fi

\tableofcontents

\input{introduction}

\input{preliminaries}

\input{language}

\input{semantics}

\input{algorithms2}

\input{implementation}

\input{rwork}

\input{conclusion}

\bibliographystyle{abbrv}
\bibliography{main}

\end{document}

%% file: line_count.tex
\newcommand{\lineTranslate}{13579\ }
\newcommand{\fileTotal}{22\ }
\newcommand{\lineTotal}{27588\ }

%% file: macros.tex
\newcommand{\true}{\mathtt{true}}
\newcommand{\false}{\mathtt{false}}
\newcommand{\synfb}{\mathtt{fdbk}}
\newcommand{\stsqltl}{\mathtt{sts2qltl}}
\newcommand{\determsts}{\mathtt{det2sts}}
\newcommand{\sts}{\mathtt{sts}}
\newcommand{\impl}{\Rightarrow}
\newcommand{\eqv}{\Leftrightarrow}

\newcommand{\stateless}{\mathtt{stateless}}
\newcommand{\determ}{\mathtt{det}}
\newcommand{\determstateless}{\mathtt{stateless\_det}}
\newcommand{\qltl}{\mathtt{qltl}}
\newcommand{\Div}{\mathtt{Div}}
\newcommand{\id}{\mathtt{Id}}

\newcommand{\synsop}{\,;}

\newcommand{\synpop}{\parallel}

\newcommand{\Next}{\mathbf{X\,}}
\newcommand{\Until}{\mathbf{~U~}}
\newcommand{\Always}{\mathbf{G\,}}
\newcommand{\Evetl}{\mathbf{F\,}}
\newcommand{\Nextvar}{\ocircle\,}
\newcommand{\Leads}{\mathbf{\,L\,}}

\newcommand{\legal}{\mathsf{legal}}

\newcommand{\sempop}{\otimes}
\newcommand{\semsop}{\circ}

\newcommand{\semfb}{\mathsf{Feedback}}
\newcommand{\synref}{\preceq}
\newcommand{\semref}{\sqsubseteq}
\newcommand{\Fusion}{\mathsf{Fusion}}
\newcommand{\sem}[1]{[\![#1]\!]}
\newcommand{\Bool}{\mathbb{B}}
\newcommand{\Nat}{\mathbb{N}}
\newcommand{\Real}{\mathbb{R}}

\newcommand{\In}[1]{\mathsf{in}(#1)}
\newcommand{\run}{\mathsf{run}}
\newcommand{\illegal}{\mathsf{illegal}}

\newcommand{\Magic}{\mathsf{Magic}}
\newcommand{\Fail}{\mathsf{Fail}}
\newcommand{\Skip}{\mathsf{Skip}}
\newcommand{\IterateOmega}{\mathsf{IterateOmega}}

\newcommand{\intype}{\Sigma_{in}}
\newcommand{\outtype}{\Sigma_{out}}
\newcommand{\fst}{\mathsf{fst}}
\newcommand{\snd}{\mathsf{snd}}
\newcommand{\Unit}{\mathsf{Unit}}

\newcommand{\symblegal}{\mathbf{legal}}
\newcommand{\serial}{\mathbf{serial}}
\newcommand{\Parallel}{\mathbf{parallel}}
\newcommand{\symbolic}{\mathbf{atomic}}
\newcommand{\OI}{\mathbf{OI}}

\newcommand{\wf}{\mathbf{wf}}
\newcommand{\symfb}{\mathbf{feedback}}
\newcommand{\deterministic}{\mathbf{determ}}
\newcommand{\decomp}{\mathbf{loop\mbox{-}free}}

%% file: abstract.tex

\begin{abstract}
The Refinement Calculus of Reactive Systems (RCRS) is a compositional formal framework for modeling and reasoning about reactive systems. RCRS provides a language which allows to describe atomic components as symbolic transition systems or QLTL formulas, and composite components formed using three primitive composition operators: serial, parallel, and feedback. The semantics of the language is given in terms of monotonic property transformers, an extension to reactive systems of monotonic predicate transformers, which have been used to give compositional semantics to sequential programs. RCRS allows to specify both safety and liveness properties. It also allows to model input-output systems which are both non-deterministic and non-input-receptive (i.e., which may reject some inputs at some points in time), and can thus be seen as a behavioral type system. RCRS provides a set of techniques for symbolic computer-aided reasoning, including compositional static analysis and verification. RCRS comes with a publicly available implementation which includes a complete formalization of the RCRS theory in the Isabelle proof assistant.
\end{abstract}

%% file: introduction.tex

\section{Introduction}
\label{sec:introduction}

This paper presents the {\em Refinement Calculus of Reactive Systems} (RCRS), a comprehensive framework for compositional modeling of and reasoning about reactive systems. 
RCRS originates from the precursor theory of {\em synchronous relational interfaces}~\cite{tripakisEMSOFT09,TripakisLHL11}, and builds upon the classic {\em refinement calculus}~\cite{backwright:98}.
A number of conference publications on RCRS exist~\cite{PreoteasaT14,DragomirPT16,PreoteasaTripakisLICS2016,DragomirPreoteasaTripakisFORTE2017,RCRS_Toolset_TACAS_2018}. 
This paper collects some of these results and extends them in significant ways. The novel contributions of this paper and relation to our previous work are presented in \S\ref{sec_contributions}.

The motivation for RCRS stems from the need for a compositional treatment of
reactive systems. Generally speaking, compositionality is a divide-and-conquer
principle. As systems grow in size, they grow in complexity. Therefore dealing
with them in a monolithic manner becomes unmanageable. Compositionality comes
to the rescue, and takes many forms. Many industrial-strength systems have
employed for many years mechanisms for compositional modeling. An example
is the Simulink tool from the Mathworks. Simulink is based on the widespread
notation of hierarchical block diagrams. Such diagrams are both intuitive,
and naturally compositional: a block can be refined into sub-blocks, sub-sub-blocks,
and so on, creating hierarchical models of arbitrary depth. This allows the user
to build large models (many thousands of blocks) while at the same time managing
their complexity (at any level of the hierarchy, only a few blocks may be visible).

But Simulink's compositionality has limitations, 
despite its hierarchical modeling approach.
Even relatively simple problems, such as the problem of modular code generation
(generating code for a block independently from context) require techniques
not always available in standard code generators~\cite{LublinermanTripakisDATE08,LublinermanTripakisPOPL09}.
Perhaps more serious, and more relevant in the context of this paper, is Simulink's lack of formal semantics, and consequent lack of rigorous analysis techniques that
can leverage the advances in the fields of computer-aided verification
and programming languages.

RCRS provides a compositional formal semantics for Simulink in particular,
and hierarchical block diagram notations in general, 
by building on well-established principles from the formal methods and
programming language domains. In particular, RCRS relies on the 
notion of {\em refinement} (and its counterpart, {\em abstraction})
which are both fundamental in system design. 
Refinement is a binary relation between components, and ideally characterizes
{\em substitutability}: the conditions under which some component can replace
another component, without compromising the behavior of the overall system.
RCRS refinement is {\em compositional} in the sense that it is preserved
by composition: if $A'$ refines $A$ and $B'$ refines $B$, then the composition
of $A'$ and $B'$ refines the composition of $A$ and $B$.

RCRS can be viewed as a refinement theory. It can
also be viewed as a {\em behavioral type system}, similar to type systems
for programming languages, but targeted to reactive systems.
By {\em behavioral}
we mean a type system that can capture not just data types of input and
output ports of components (bool, int, etc.), but also complete specifications
of the behavior of those components.
As discussed more extensively in~\cite{TripakisLHL11}, such a behavioral
type system has advantages over a full-blown verification system, as it
is more lightweight: for instance, it allows type checking, which does not
require the user to provide a formal specification of the properties that
the model must satisfy, the model must simply {\em type-check}.

To achieve this, it is essential, as argued in~\cite{TripakisLHL11}, 
 for the framework to be able to express
{\em non-input-receptive} (also called {\em non-input-enabled} or
{\em non-input-complete}) components, i.e., components that reject some
input values.
RCRS allows this. For example, a square-root component can be described
in RCRS alternatively as: (1) a non-input-receptive component $C_1$ with input-output
{\em contract} $x \ge 0 \land y=\sqrt{x}$ (where $x,y$ are the input and output
variables, respectively); or (2) an input-receptive component $C_2$ with contract
$x \ge 0 \to y=\sqrt{x}$.
Simple examples of type-checking in the RCRS context are the following:
Connecting the non-input-receptive square-root component $C_1$ to a component
which outputs $x=-1$ results in a type error ({\em incompatibility}).
Connecting $C_1$ to a non-deterministic component which outputs an arbitrary
value for $x$ (this can be specified by the formula/contract $\mathit{true}$) also
results in a type error. 
Yet, in both of the above cases $C_2$ will output a non-deterministically 
chosen value, even though the input constraint is not satisfied.

RCRS allows both non-input-receptive and non-deterministic components.
This combination results in a {\em game-theoretic} interpretation of the 
composition operators, like in {\em interface theories}~\cite{AlfaroHenzingerFSE01,TripakisLHL11}. Refinement also becomes game-theoretic, as in
{\em alternating refinement}~\cite{AlternatingCONCUR98}.
Game-theoretic composition can be used for an interesting form of type
inference. For example, if we connect the non-input-receptive square-root component $C_1$ above to a non-deterministic component $C_3$ with input-output contract $x\ge u+1$ (where $x$ is the output of $C_3$, and $u$ its input), and apply
the (game-theoretic) serial composition of RCRS, we obtain the condition
$u\ge -1$ on the external input of the overall composition. The constraint
$u\ge -1$ represents the weakest condition on $u$ which ensures compatibility
of the connected components.

In a nutshell, RCRS consists of the following elements:
\begin{enumerate}
\item A modeling language (syntax), which allows to describe {\em atomic} components,
and {\em composite} components formed by a small
number of primitive composition operators (serial, parallel, and feedback).
The language is described in \S\ref{sec:syntax}.
\item A formal semantics, presented in \S\ref{sec:semantics}.
Component semantics are defined in terms of {\em monotonic
property transformers} (MPTs). MPTs are extensions of monotonic {\em predicate}
transformers used in theories of programming languages, and in particular
in refinement calculus~\cite{backwright:98}.
Predicate transformers transform sets of {\em post-states} (states reached
by the program after its computation) into sets of {\em pre-states} (states
where the program begins). Property transformers transform sets of a
component's {\em output traces} (infinite sequences of output values) into 
sets of {\em input traces} (infinite sequences of input values). 
Using this semantics we can express both safety and liveness properties.
\item A set of symbolic reasoning techniques, described in \S\ref{sec:algorithms}. 
In particular, RCRS offers techniques to
\begin{itemize}
\item compute the symbolic representation of a
composite component from the symbolic representations of its sub-components;
\item simplify composite components into atomic components;
\item reduce checking refinement between two components
to checking satisfiability of certain logical formulas;
\item reduce input-receptiveness and compatibility checks to satisfiability;
\item compute the legal inputs of a component symbolically.
\end{itemize}
We note that these techniques are for the most part {\em logic-agnostic},
in the sense that they do not depend on the particular logic used to
represent components.

\item A toolset, described briefly in \S\ref{sec_implementation}.
The toolset consists mainly of:
\begin{itemize}
\item a full implementation of the RCRS theory in the
Isabelle proof assistant~\cite{NipkowPW02};
\item a translator of Simulink diagrams into RCRS code.
\end{itemize}
Our implementation is open-source and publicly available from \url{http://rcrs.cs.aalto.fi}.
\end{enumerate}

\subsection{Novel contributions of this paper and relation to our prior work}
\label{sec_contributions}

Several of the ideas behind RCRS originated in the theory of synchronous relational interfaces~\cite{tripakisEMSOFT09,TripakisLHL11}. The main novel contributions of RCRS w.r.t. that theory are: (1) RCRS is based on the semantic foundation of monotonic property transformers, whereas relational interfaces are founded on relations; (2) RCRS can handle liveness properties, whereas relational interfaces can only handle safety; (3) RCRS has been completely formalized and most results reported in this and other RCRS papers have been proven in the Isabelle proof assistant; (4) RCRS comes with a publicly available toolset (\url{http://rcrs.cs.aalto.fi}) which includes the Isabelle formalization, a Translator of Simulink hierarchical block diagrams~\cite{DragomirPT16,RCRS_Toolset_TACAS_2018}, and a Formal Analyzer which performs, among other functions, compatibility checking, refinement checking, and automatic simplification of RCRS contracts.

RCRS was introduced in~\cite{PreoteasaT14}, which focuses on monotonic
property transformers as a means to extend relational interfaces with liveness
properties. \cite{PreoteasaT14} covers serial composition, but not parallel
nor feedback. It also does not cover symbolic reasoning nor the RCRS implementation.
Feedback is considered in~\cite{PreoteasaTripakisLICS2016}, whose aim is in
particular to study {\em instantaneous feedback} 
for non-deterministic and non-input-receptive systems.
The study of instantaneous feedback is an interesting problem, but beyond
the scope of the current paper. 
In this paper we consider non-instantaneous feedback, i.e., feedback
for systems without same-step cyclic dependencies (no {\em algebraic loops}).

\cite{DragomirPT16} presents part of the RCRS implementation, focusing on the translation
of Simulink (and hierarchical block diagrams in general) into an algebra
of components with three composition primitives, serial, parallel, and
feedback, like RCRS. As it turns out, there is not a unique way to translate
a graphical notation like Simulink into an algebraic formalism like RCRS.
The problem of how exactly to do it and what are the trade-offs is an
interesting one, but beyond the scope of the current paper. This problem
is studied in depth in~\cite{DragomirPT16} which proposes three different
translation strategies and evaluates their pros and cons. \cite{DragomirPT16} 
leaves open the question whether the results obtained by the different
translations are equivalent. \cite{PDT:2016}
settles this question, by proving that a class of translations, including
the ones proposed in~\cite{DragomirPT16} are semantically equivalent
for any input block diagram.
\cite{DragomirPreoteasaTripakisFORTE2017} also concerns the RCRS implementation,
discussing solutions to subtle typing problems that arise when translating
Simulink diagrams into RCRS/Isabelle code.

In summary, the current paper does not cover the topics covered in~\cite{DragomirPT16,PDT:2016,DragomirPreoteasaTripakisFORTE2017,PreoteasaTripakisLICS2016} 
and can be seen as a significantly revised and extended version of~\cite{PreoteasaT14}. The main novel contributions with respect to~\cite{PreoteasaT14} are the following:
(1) a language of components (\S\ref{sec:syntax});
(2) a revised MPT semantics (\S\ref{sec:semantics}), including in particular 
novel operators for feedback (\S\ref{sec_novel_semantical_operators_for_feedback}) and a classification of MPT subclasses (\S\ref{subsec:mpt_classes});
(3) a new section of symbolic reasoning (\S\ref{sec:algorithms}).

%% file: preliminaries.tex

\section{Preliminaries}
\label{sec:prelim}

\paragraph{Sets, types.}
We use capital letters $X$, $Y$, $\Sigma$, $\ldots$ to denote types or sets, and small letters to denote 
elements of these types $x \in X$, $y \in Y$, etc. 
We denote by $\Bool$ the type of Boolean values $\mathsf{true}$ and $\mathsf{false}$. 
We use $\land$, $\lor$, $\impl$, and $\neg$ for the Boolean operations. The type of natural 
numbers is denoted by $\Nat$, while the type of real numbers is denoted by $\Real$.
The $\Unit$ type contains a single element denoted $()$.

\paragraph{Cartesian product.}
For types $X$ and $Y$, $X \times Y$ is the Cartesian product of $X$ and $Y$, and 
if $x \in X$ and $y \in Y$, then $(x, y)$ is a tuple from $X \times Y$.
The empty Cartesian product is $\Unit$.
We assume that we have only flat products $X_1 \times \ldots \times X_n$, and then we have 
$$
(X_1 \times \ldots \times X_n)\times (Y_1 \times \ldots \times Y_m) = 
   X_1 \times \ldots \times X_n \times Y_1 \times \ldots \times Y_m
$$

\paragraph{Functions.} If $X$ and $Y$ are types, $X \to Y$ denotes the type of \textit{functions} from $X$ to $Y$. The function type constructor associates to the right 
(e.g., $X \to Y \to Z = X \to (Y \to Z) $) and the function interpretation associates to the left (e.g., $f(x)(y) = (f(x))(y)$). In order to construct functions we use 
lambda notation, e.g., $(\lambda x, y: x + y + 1): \Nat \to \Nat \to \Nat$. Similarly, we can have tuples in the definition of functions, 
e.g., $(\lambda (x, y): x + y + 2): (\Nat \times \Nat) \to \Nat$. The composition of two functions 
$f: X \to Y$ and $g : Y \to Z$, is a function denoted $g \circ f: X \to Z$, where $(g \circ f) (x) = g(f(x))$.

\paragraph{Predicates.} A {\em predicate} is a function returning Boolean values, e.g., $p: X \to Y \to \Bool$, $p(x)(y) = (x = y)$. We define the {\em smallest predicate} 
$\bot : X \to \Bool$ where $\bot(x) = \mathsf{false}$ for all $x\in X$. The {\em greatest predicate} is $\top : X \to \Bool$, with $\top(x) = \mathsf{true}$ for all $x\in X$.
We will often interpret predicates as sets. A predicate $p : X\to\Bool$ can
be viewed as the set of all $x\in X$ such that $p(x) = \mathsf{true}$.
For example, viewing two predicates $p,q$ as sets, we can write $p\subseteq q$,
meaning that for all $x$, $p(x) \impl q(x)$.

\paragraph{Relations.} A {\em relation} is a predicate with at least two arguments, e.g., $r: X \to Y \to \Bool$. For such a relation $r$, 
we denote by $\In{r} : X \to \Bool$ the predicate $\In{r}(x) = (\exists y: r(x)(y))$. If the relation $r$ has more than two arguments, then we define $\In{r}$ similarly by quantifying over the last argument.

We extend point-wise all operations on Booleans to operations on predicates and relations. For example, if $r, r' : X \to Y \to \Bool$ are two relations, then $r \land r'$ and $r \lor r'$ are 
the relations given by $(r \land r')(x)(y) = r(x)(y) \land r'(x)(y)$
and $(r \lor r')(x)(y) = r(x)(y) \lor r'(x)(y)$.
We also introduce the order on relations $r \subseteq r' = (\forall x, y: r(x)(y) \impl r'(x)(y))$.

The composition of two relations $r: X \to Y \to \Bool$ and $r': Y \to Z \to \Bool$ is a relation $(r \semsop r'): X \to Z \to \Bool$, 
where $(r \semsop r')(x)(z) = (\exists y: r(x)(y) \land r'(y)(z))$.

\delete{
\paragraph{Variables, Types, and \blue{Valuations}.} 
Let $X$ be a set of {\em variables}. 
We assume that each variable $x\in X$ is implicitly typed, i.e., that it 
comes with a set $V_x$ over which the values of $x$ range.
$V_x$ is also called the type of $x$.
Let $V_X = \bigcup_{x\in X}V_x$.
A {\em valuation} over $X$ is a function $s : X \to V_X$ such that
$s(x)\in V_x$ for all $x\in X$.
Thus, a valuation over $X$ is a total function that assigns a value to every 
variable in $X$.
The set of all valuations over $X$ is denoted $\Sigma_X$, i.e.,
$\Sigma_X = (X\to V_X)$.
If $s$ is a valuation over $X$ and $x\in X$, then
$s[x:=a]$ is the new valuation obtained from $s$ by changing the value of variable $x$ to $a$, where $a$ is assumed to be a value in $V_x$.
\red{possible notation clash: $s$ might later be a state variable? in that case use $v$ for valuation}
}

\paragraph{Infinite sequences.} If $\Sigma$ is a type, 
then $\Sigma^{\omega} = (\Nat \to \Sigma)$ is the set of all {\em infinite sequences} over $\Sigma$, also called {\em traces}.
For a trace $\sigma \in \Sigma^{\omega}$, let $\sigma_i = \sigma(i)$ be the $i$-th element in the trace.
Let $\sigma^{i} \in \Sigma^{\omega}$ denote the suffix of $\sigma$ starting
from the $i$-th step, 
i.e., $\sigma^{i} = \sigma_i \sigma_{i+1} \cdots$.
We often view a pair of traces $(\sigma, \sigma')\in \Sigma^\omega \times \Sigma'^\omega$ as being also 
a trace of pairs $(\lambda i: (\sigma_{i}, \sigma'_{i})) \in (\Sigma \times \Sigma')^\omega$. 

\paragraph{Properties.} A \textit{property} is a predicate $p$ over a set of infinite sequences. 
Formally, $p \in (\Sigma^{\omega} \to \Bool)$.
Just like any other predicate, a property can also be viewed as a set.
In particular, a property can be viewed as a set of traces.

%% file: language.tex

\section{Language}
\label{sec:syntax}

\subsection{An Algebra of Components}
\label{subsec:component_grammar}

We model systems using a simple language of {\em components}.
The grammar of the language is as follows:
\begin{center}
\begin{tabular}{rcl}
component & ::= & atomic\_component $|$ composite\_component \\
atomic\_component & ::= & STS\_component $|$ QLTL\_component \\
STS\_component & ::= & GEN\_STS\_component $|$ STATELESS\_STS\_component \\ 
		& & $|$ DET\_STS\_component $|$ DET\_STATELESS\_STS\_component \\
composite\_component & ::= & component $\synsop$ component 
		$|$ component $\synpop$ component
		$|$ $\synfb$(component)
\end{tabular}
\end{center}

The elements of the above grammar are defined in the remaining of this
section, where examples are also given to illustrate the language.
In a nutshell, the language contains {\em atomic} components of two kinds:
atomic components defined as {\em symbolic transition systems}
(STS\_component), and atomic components defined as QLTL formulas
over input and output variables (QLTL\_component). 

STS components are split in four categories: general STS components,
stateless STS components, deterministic STS components, and deterministic
stateless STS components. Semantically, the general STS components subsume
all the other more specialized STS components, but we introduce the specialized 
syntax because symbolic compositions of less general components become simpler,
as we shall explain in the sequel (see \S\ref{sec:algorithms}).

Also, as it turns out, atomic components of our framework form a lattice,
shown in Fig.~\ref{fig:basic}, from the more specialized ones, namely,
deterministic stateless STS components, to the more general ones, namely
QLTL components.
The full definition of this lattice will become apparent once we provide
a symbolic transformation of STS to QLTL components, in \S\ref{sec_sts2qltl}.

\begin{figure}[!t]
	\centering
	\subfloat[Serial composition: $C\synsop C'$]{
		\begin{tikzpicture}
		  \node[draw, minimum height = 4ex] (a) {$C$};
		  \node[left=4ex of a.west] (px) {};
		  \node[shape=coordinate, right = 3ex of a.east] (py) {};
		  \node[draw, minimum height = 4ex, right = 8ex of a] (b) {$C'$};
		  \node[right=4ex of b.east] (pv) {};
		  \node[shape=coordinate, left = 3ex of b.west] (pu) {};
		  \draw[-latex'] (px) --++ (2ex, 0) node[anchor=south] {$x$} -- (px -| a.west);
		  \draw[-latex'] (b.east) --++ (3ex, 0) node[anchor=south] {$z$} -- (b.east -| pv);
		  \draw[-latex'] (a.east) --++ (1ex, 0) node[anchor=south] {$y$} -- (py);
		  \draw[-latex'] (pu) --++ (2ex, 0) node[anchor=south] {$y$} -- (b.west);
		  \draw[dashed] (py) -- (pu);
		  \node[shape=coordinate, above left = 0.1ex and 0.7ex of a](pa){};
		  \node[shape=coordinate, below right = 0.1ex and 0.7ex of b, inner sep = 0](pb){};
		  \node[draw, style=dotted, thick, inner ysep = 2ex, fit = (a)(b)(pa)(pb)](main){};
		\end{tikzpicture}
		\label{fig:serial_comp}
	} 
	\qquad
	\subfloat[Parallel composition: $C\synpop C'$]{
		\hspace*{30pt}
		\begin{tikzpicture}
		  \node[draw, minimum height = 4ex] (a) {$C$};
		  \node[draw, minimum height = 4ex, below = 3ex of a] (b) {$C'$};
		  \node[left=4ex of a.west] (x) {};
		  \node[right=4ex of a.east] (y) {};
		  \node[left=4ex of b.west] (u) {};
		  \node[right=4ex of b.east] (v) {};
		  \draw[-latex'] (x) --++ (2ex, 0) node[anchor=south] {$x$} -- (x -| a.west);
		  \draw[-latex'] (u) --++ (2ex, 0) node[anchor=south] {$u$} -- (u -| b.west);
		  \draw[-latex'] (a.east) --++ (3ex, 0) node[anchor=south] {$y$} -- (a.east -| y);
		  \draw[-latex'] (b.east) --++ (3ex, 0) node[anchor=south] {$v$} -- (b.east -| v.west);
		  \node[shape=coordinate, left = 0.7ex of b](pa){};
		  \node[shape=coordinate, right = 0.7ex of b, inner sep = 0](pb){};
		  \node[draw, style=dotted, thick, fit = (a)(b)(pa)(pb)](main){};
		\end{tikzpicture}
		\hspace*{10pt}
		\label{fig:parallel_comp}
	}
	\qquad
	\subfloat[Feedback composition: $\synfb(C)$]{
		\hspace*{4pt}
		\begin{tikzpicture}
		  \node[draw, minimum height = 10ex, minimum width=3ex] (a) {$C$};
		  \node[left = 1ex of a.north west, anchor = north east](x){$x_1$};
		  \draw[-latex'](x.south west) -- (x.south west-|a.west);
		  \node[right = 1ex of a.north east, anchor = north west](y){$y_1$};
		  \draw[latex'-](y.south east) -- (x.south west-|a.east);
		  \draw[dotted](y.south east) -- ++(0,4ex) coordinate(A) -- (A -| x.west) -- (x.south west);
		  \coordinate[left = 4.5ex of x](X);	  
		  \node[below = 2ex of X](xa){$x_2$};
		  \node[below = -1ex of xa](xb){$\vdots$};
		  \draw[-latex'](xa) -- (xa -| a.west);

		  \coordinate[right = 5ex of y](Y);	  
		  \node[below = 2ex of Y](ya){$y_2$};
		  \node[below = -1ex of ya](yb){$\vdots$};
		  \draw[-latex'](ya -| a.east) -- (ya);

		  \node[draw, style=dotted,inner ysep = 2ex, thick, fit = (x)(y)(a)]{};

		\end{tikzpicture}
		\hspace*{4pt}
		\label{fig:fb_comp}
	}
	\caption{The three composition operators of RCRS.}
	\label{fig:comp_ops}
\end{figure}
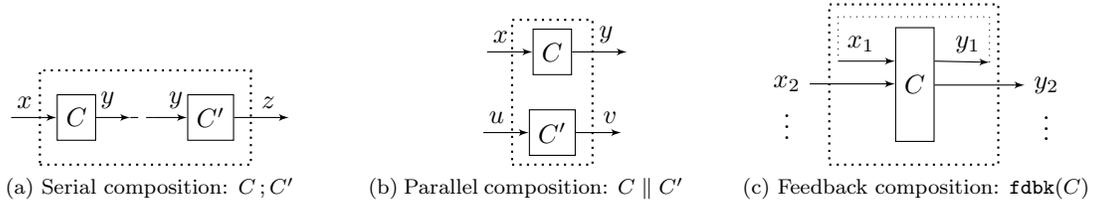

Apart from atomic components, the language also allows to form
{\em composite} components, by composing (atomic or other composite) components
via three composition operators: serial $\synsop$,
parallel $\synpop$, and feedback $\synfb$, as depicted in Fig.~\ref{fig:comp_ops}.
The serial composition of two components $C,C'$ is formed by connecting the output(s) of $C$ to the input(s) of $C'$. 
Their parallel composition is formed by ``stacking'' the two components on top of each other without forming any new connections.
The feedback of a component $C$ is obtained by connecting the first output 
of $C$ to its first input.

Our language is inspired by graphical notations such as Simulink, and
hierarchical block diagrams in general.
But our language is textual, not graphical. 
An interesting question is how to translate a graphical block diagram into 
a term in our algebra. We will not address this question here, as the issue is 
quite involved. We refer the reader to~\cite{DragomirPT16}, which includes
an extensive discussion on this topic.
Suffice it to say here that there are generally many possible translations of a
graphical diagram into a term in our algebra (or generally any algebra
that contains primitive serial, parallel, and feedback composition operators).
These translations achieve different tradeoffs in terms of size, readability,
computational properties, and so on. See~\cite{DragomirPT16} for details.

\begin{figure}[!h]
  \centering
\ifdefined\arxivversion
  \includegraphics[scale=0.7]{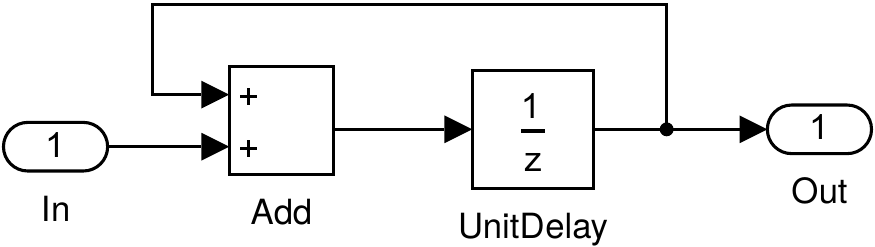}
\else
  \includegraphics[scale=0.7]{../figures/diag}
\fi
  \caption{A Simulink diagram modeling the 1-step delayed sum of its input \texttt{In}. Each atomic block as well as the entire system can be formalized as STS components (see \S\ref{subsec:sts}). The entire system can also be formalized as a composite component (see below).}
  \label{fig:diagram}
\end{figure}

As an example, consider the Simulink diagram shown in Fig.~\ref{fig:diagram}.
This diagram can be represented in our language as a composite component
$\mathtt{Sum}$ defined as 
$$\mathtt{Sum} = \synfb(\mathtt{Add} \synsop \mathtt{UnitDelay} \synsop \mathtt{Split})$$
where $\mathtt{Add}$, $\mathtt{UnitDelay}$, and $\mathtt{Split}$ are atomic components (for a definition of these atomic components see \S\ref{subsec:sts}).
Here $\mathtt{Split}$ models the ``fan-out'' element in the Simulink diagram (black bullet) where the output wire of $\mathtt{UnitDelay}$ splits in two wires going to $\mathtt{Out}$ and back to $\mathtt{Add}$.\footnote{
Note that the Simulink input and output ports $\mathtt{In}$ and $\mathtt{Out}$ 
are not explicitly represented in $\mathtt{Sum}$. They are represented
implicitly: $\mathtt{In}$ corresponds to the second input of $\mathtt{Add}$,
which carries over as the unique external input of $\mathtt{Sum}$ (thus,
$\mathtt{Sum}$ is an ``open system'' in the sense that it has open,
unconnected, inputs); $\mathtt{Out}$ corresponds to the second output
of $\mathtt{Split}$, which carries over as the unique external output
of $\mathtt{Sum}$.
} 

\subsection{Symbolic Transition System Components}
\label{subsec:sts}

We introduce all four categories of STS components and at the same time
 provide syntactic mappings from specialized STS
components to general STS components.

\subsubsection{General STS Components} 
\label{subsubsec:sts}

A {\em general symbolic transition system component} 
(general STS component) is a transition system described
symbolically, with Boolean expressions over input, output, state, and next state variables
defining the initial states and the transition relation.
When we say ``Boolean expression'' (here and in the definitions that follow)
we mean an expression of type $\Bool$, in
some arbitrary logic, not necessarily restricted to propositional logic.
For example, if $x$ is a variable of numerical type, then $x>0$ is a Boolean
expression.
In the definition that follows, $s'$ denotes the {\em primed}, or {\em next
state} variable, corresponding to the {\em current state} variable $s$.
Both can be vectors of variables.
For example, if $s=(s_1,s_2)$ then $s'=(s_1',s_2')$. We assume that $s'$ has the same type as $s$.

\begin{definition}[STS component]
  A {\em (general) STS component} is a tuple 
  $$\sts(x:\Sigma_x, y:\Sigma_y, s:\Sigma_s, \mathit{init\_exp}:\Bool, \mathit{trs\_exp}:\Bool)$$
  where $x,y,s$ are input, output and state variables (or tuples of variables)
  of types $\Sigma_x, \Sigma_y,\Sigma_s$, respectively,
  $\mathit{init\_exp}$ is a Boolean expression on $s$ (in some logic),
  and $\mathit{trs\_exp}$ is a Boolean expression on $x,y,s,s'$ (in some logic).
\end{definition}

Intuitively, an STS component 
is a non-deterministic system which for an infinite
input sequence $\sigma_x\in\Sigma_x^\omega$ produces as output an infinite sequence $\sigma_y \in \Sigma_y^\omega$.
The system starts non-deterministically at some state $\sigma_s(0)$ satisfying $\mathit{init\_exp}$.
Given first input $\sigma_x(0)$, the system non-deterministically computes output
$\sigma_y(0)$ and next state $\sigma_s(1)$ such that $\mathit{trs\_exp}$ holds
(if no such values exist, then the input $\sigma_x(0)$ is {\em illegal}, as
discussed in more detail below).
Next, it uses the following input $\sigma_x(1)$ and state $\sigma_s(1)$ to compute $\sigma_y(1)$ and $\sigma_s(2)$, and so on.

We will sometimes use the term {\em contract} to refer to the expression
$\mathit{trs\_exp}$.
Indeed, $\mathit{trs\_exp}$ can be seen as specifying a contract between
the component and its environment, in the following sense. At each step
in the computation, the environment must provide input values that do
not immediately violate the contract, i.e., for which we can find values
for the next state and output variables to satisfy $\mathit{trs\_exp}$. 
Then, it is the responsibility of the component to find such values,
otherwise it is the component's ``fault'' if the contract is violated.
This game-theoretic interpretation is similar in spirit with the classic
refinement calculus for sequential programs~\cite{backwright:98}.

We use $\Sigma_x$, $\Sigma_y$, $\Sigma_s$ in the definition above to emphasize the
types of the input, output and the state, and the fact that, when composing components, 
the types should match. However, in practice we often omit
the types, unless they are required to unambiguously specify a component.
Also note that the definition does not fix the logic used for the expressions $\mathit{init\_exp}$ and $\mathit{trs\_exp}$. 
Indeed, our theory and results are independent from the choice of this logic.
The choice of logic matters for algorithmic complexity and decidability. We will return to this point in \S\ref{sec:algorithms}.
Finally, for simplicity, we often view the formulas $\mathit{init\_exp}$ and $\mathit{trs\_exp}$ as semantic objects, namely, as predicates. 
Adopting this view, $\mathit{init\_exp}$ becomes the predicate $init:\Sigma_s\to\Bool$, and
$\mathit{trs\_exp}$ the predicate $\mathit{trs\_exp} : \Sigma_s\to\Sigma_x\to\Sigma_s\to\Sigma_y\to\Bool$.
Equivalently, $\mathit{trs\_exp}$ can be interpreted as a relation $\mathit{trs\_exp} : (\Sigma_s\times\Sigma_x)\to(\Sigma_{s}\times\Sigma_y)\to\Bool$.

Throughout this paper we assume that $\mathit{init\_exp}$ is satisfiable,
meaning that there is at least one valid initial state.

\paragraph{Examples.}

In the examples provided in this paper, we often 
specify systems that have tuples as input, state and output variables,
in different equivalent ways.  For example, we can introduce a general STS component with two inputs as $\sts((n:\Nat,x:\Real), s:\Real, y:\Real, s > 0 , s' > s \land  y + s = x ^ n)$, 
but also as $\sts((n,x):\Nat\times \Real, s:\Real, y:\Real, s > 0 , s' > s \land  y + s = x ^ n)$, or $\sts(z:\Nat\times\Real, y:\Real,s:\Real, s > 0 , s' > s \land  y + s= \snd(z) ^ {\fst(z)})$,
where $\fst$ and $\snd$ return the first and second elements of a pair.

Let us model a system that at every step $i$ outputs the input received at previous step $i-1$ (assume that the initial output value is $0$).
This corresponds to Simulink's commonly used $\mathtt{UnitDelay}$ block,
which is also modeled in the diagram of Fig.~\ref{fig:diagram}. This block 
can be represented by an STS component, 
where a state variable $s$ is needed to store the input at moment $i$ such that it can be used at the next step. 
We formally define this component as 
$$\mathtt{UnitDelay} = \sts(x, y, s, s=0,  y=s \land s'=x).$$
 We use first-order logic to define $\mathit{init\_exp}$ and $\mathit{trs\_exp}$.
Here $\mathit{init\_exp}$ is $s = 0$, which initializes the state variable $s$ with the value $0$.
The $\mathit{trs\_exp}$ is $y = s \land s'=x$, that is at moment $i$ the current state variable $s$ which stores the input $x$ received at moment $i-1$ is outputed and its value is updated. 

As another example, consider again the composite component $\mathtt{Sum}$ 
modeling the diagram of Fig.~\ref{fig:diagram}.
$\mathtt{Sum}$ could also be defined as an atomic STS component:
$$\mathtt{Sum} = \sts(x, y, s, s=0, y=s \land s'=s+x).$$
In \S\ref{sec:algorithms} we will show how we can automatically and symbolically
simplify composite component terms such as
$\synfb(\mathtt{Add} \synsop \mathtt{UnitDelay} \synsop \mathtt{Split})$,
to obtain syntactic representations of atomic components
such as the one above.

These examples illustrate systems coming from Simulink models. However, our language is more general, and able to accommodate the description of other systems, such as state machines 
\`a la nuXmv \cite{CavadaCDGMMMRT14}, or input/output automata \cite{LynchTuttle89}.
In fact, both $\mathtt{UnitDelay}$ and $\mathtt{Sum}$ are deterministic, so they could also be defined as deterministic STS components, as we will see below.
Our language can capture non-deterministic systems easily.
An example of a non-deterministic STS component is the following: 
$$C = \sts(x, y, s, s = 0, x + s \leq y).$$
For an input sequence $\sigma_x \in \Nat^\omega$, $C$ outputs a non-deterministically chosen sequence $\sigma_y$ such that the transition expression $x + s \leq y$ is satisfied. 
Since there is no formula in the transition expression tackling the next state variable, this is updated also non-deterministically with values from $\Nat$. 

Our language can also capture {\em non-input-receptive} systems, that is,
systems which disallow some input values as {\em illegal}.
For instance, a component performing division, but disallowing division by zero,
 can be specified as follows:
$$\Div = \sts((x,y),z,(),\true,y\ne 0\land z=\dfrac{x}{y}).$$ 
Note that $\Div$ has an empty tuple of state variables, $s=()$.
Such components are called {\em stateless}, and are introduced in the sequel.

Even though RCRS is primarily a discrete-time framework, we have used it
to model and verify continuous-time systems such as those modeled in
Simulink (see \S\ref{sec_implementation}). We do this by discretizing
time using a time step parameter $\Delta t>0$ and applying Euler numerical
integration. Then, we can model Simulink's {\em Integrator} block in
RCRS as an STS component parameterized by $\Delta t$:

\begin{eqnarray*}
\mathtt{Integrator}_{\Delta t} &=& \sts\big(x, y, s, s=0, y=s \land s'=s+x\cdot\Delta t \big)
\end{eqnarray*}

More complex dynamical system blocks can be modeled in a similar fashion.
For instance, Simulink's {\em Transfer Fcn} block, with transfer function
$$
\frac{s^2+2}{0.5s^2 + 2s + 1}
$$
can be modeled in RCRS as the following STS component parameterized by $\Delta t$:

\begin{eqnarray*}
\mathtt{TransferFcn}_{\Delta t} &=& \sts\big(x, y, (s_1,s_2), s_1=0\land s_2=0,
	trs) \\
\mbox{ where } \qquad trs &=& (y = -8\cdot s_1 + 2\cdot x)\ 
  \land\\
			 & & \qquad (s_1' = s_1 + (-4\cdot s_1 -2\cdot s_2 + x) \cdot\Delta t) \ \land  \\
			 & & \qquad (s_2' = s_2 + s_1\cdot\Delta t)
\end{eqnarray*}


\subsubsection{Variable Name Scope}
\label{subsubsec:name_scope}

We remark that variable names in the definition of
atomic components are {\em local}. This holds for all atomic components
in the language of RCRS (including STS and QLTL components, defined in
the sequel). 
This means that if we replace a variable with another one in an atomic component, then
we obtain a semantically equivalent component.
For example, the two STS components below are equivalent
(the semantical equivalence symbol $\equiv$ will be defined formally
in Def.~\ref{def_sem_equiv_refin}, once we define the semantics):
$$
\sts((x,y),z,s,s > 0, z > s + x + y) \equiv \sts((u,v),w,t,t > 0, w > t + u + v)
$$

\subsubsection{Stateless STS Components} 
\label{subsubsec:stateless_sts}

A special STS component is one that has no state variables:

\begin{definition}[Stateless STS component]
A {\em stateless STS component} is a tuple
$$C = \stateless(x:\Sigma_x,y:\Sigma_y,\mathit{io\_exp}:\Bool)$$
where $x,y$ are the input and output variables,
and $\mathit{io\_exp}$ is a Boolean expression on $x$ and $y$.
Stateless STS components are special cases of general STS components,
as defined by the mapping $\stateless 2 \sts$:
$$
\stateless 2 \sts(C) = \sts(x, y, (), \true, \mathit{io\_exp}).
$$
\end{definition}

\paragraph{Examples.}
A trivial stateless STS component is the one that simply
transfers its input to its output (i.e., a ``wire'').
We denote such a component by $\id$, and we formalize it as 
$$\id = \stateless(x, y, y=x).$$

Another simple example is a component with no inputs and a single output,
which always outputs a constant value $c$ (of some type). This can be 
formalized as the following component parameterized by $c$:
$$\mathtt{Const}_c = \stateless((), y, y = c ).$$

Component $\mathtt{Add}$ from Fig.~\ref{fig:diagram}, which outputs the sum of its two inputs, can be modeled as a stateless STS component:
$$\mathtt{Add} = \stateless((x,y), z, z = x + y).$$

Component $\mathtt{Split}$ from Fig.~\ref{fig:diagram} can also be
modeled as a stateless STS component:
$$\mathtt{Split}=\stateless(x, (y,z), y=x \land z=x).$$

The $\Div$ component introduced above is stateless, and therefore can be also
 specified as follows:
$$\Div = \stateless\big((x,y),z,y\ne 0\land z=\dfrac{x}{y}\big).$$ 

The above examples are not only stateless, but also deterministic.
We introduce deterministic STS components next.

\subsubsection{Deterministic STS Components}
\label{subsubsec:deterministic_sts}

Deterministic STS components are those which, for given current state and input,
have at most one output and next state.
Syntactically, they are introduced as follows:

\begin{definition}[Deterministic STS component]
	A {\em deterministic STS component} is a tuple 
	$$\determ(x:\Sigma_x, s:\Sigma_s, a:\Sigma_s, \mathit{inpt\_exp}:\Bool,\mathit{next\_exp}:\Sigma_s, \mathit{out\_exp}:\Sigma_y)$$
	where $x,s$ are the input and state variables,
	$a\in\Sigma_s$ is the initial value of the state variable, 
	$\mathit{inpt\_exp}$ is a Boolean expression on $s$ and $x$ 
	defining the legal inputs,
	$\mathit{next\_exp}$ is an expression of type $\Sigma_s$ on $x$ and $s$ defining the next state,
	and $\mathit{out\_exp}$ is an expression of type $\Sigma_y$ 
	on $x$ and $s$
	defining the output.
Deterministic STS components are special cases of general STS components,
as defined by the mapping $\determ 2\sts$:
	$$
		\determ 2\sts(C) = (x,y,s, s = a ,  \mathit{inpt\_exp} \land s' = \mathit{next\_exp}
		\land y = \mathit{out\_exp})
	$$
	where $y$ is a new variable name (or tuple of new variable names)
	of type $\Sigma_y$.
\end{definition}

Note that a deterministic STS component has a separate expression
$\mathit{inpt\_exp}$ to define legal inputs. A separate such expression
is not needed for general STS components, where the conditions for legal
inputs are part of the expression $\mathit{trs\_exp}$.
For example, compare the definition of $\mathtt{Div}$ as a general STS above,
and as a stateless deterministic STS below (see~\S\ref{subsubsec:stateless_deterministic_sts}).

\paragraph{Examples.}

As mentioned above, all three components, $\mathtt{UnitDelay}$, $\mathtt{Add}$,
and $\mathtt{Split}$ from Fig.~\ref{fig:diagram}, as well as $\mathtt{Div}$ 
and $\mathtt{Const}$, are deterministic. They could
therefore be specified in our language as deterministic STS components,
instead of general STS components:
\begin{eqnarray*}
\mathtt{UnitDelay} &=& \determ\big(x, s, 0, \true, x, s \big) \\
\mathtt{Const}_c &=& \determ\big((),(),(), \true, (), c\big) \\
\mathtt{Add} &=& \determ\big((x,y),(),(), \true, (), x + y\big) \\
\mathtt{Split}&=& \determ\big(x,(),(), \true, (), (x,x)\big) \\
\mathtt{Div}&=& \determ\big((x,y),(),(),y\not=0, (), \dfrac{x}{y}\big)
\end{eqnarray*}

The component $\mathtt{Sum}$ modeling the entire system is also deterministic,
and could be defined as a deterministic STS component:
$$\mathtt{Sum} =\determ(x, s, 0, \true, s+x, s).$$

Note that these alternative specifications for each of those components,
although syntactically distinct, will turn out to be semantically equivalent
by definition, when we introduce the semantics of our language, in \S\ref{sec:semantics}.

\subsubsection{Stateless Deterministic STS Components}
\label{subsubsec:stateless_deterministic_sts}

STS components which are both deterministic and stateless can be specified
as follows:

\begin{definition}[Stateless deterministic STS component]
A {\em stateless deterministic STS component} is a tuple 
$$C = \determstateless(x:\Sigma_x, \mathit{inpt\_exp}:\Bool, \mathit{out\_exp}:\Sigma_y)$$
 where $x$ is the input variable,
$\mathit{inpt\_exp}$ is a Boolean expression on $x$ defining the legal
inputs, and $\mathit{out\_exp}$ is an expression of type $\Sigma_y$ on x
defining the output.
Stateless deterministic STS components are special cases of both
deterministic STS components, and of stateless STS components,
as defined by the mappings 
\begin{eqnarray*}
\determstateless 2 \determ(C) & = & \determ(x,(),(),\mathit{inpt\_exp},(),\mathit{out\_exp}) \\
\determstateless 2 \stateless(C) & = & \stateless(x,y,\mathit{inpt\_exp} \land y = \mathit{out\_exp})
\end{eqnarray*}
where $y$ is a new variable name or a tuple of new variable names.
\end{definition}

\paragraph{Examples.}

Many of the examples introduced above are both deterministic and stateless. 
They could be specified as follows:
$$
  \begin{array}{lll}
  \id &=& \determstateless\big(x, \true, x\big)\\[1ex]
  \mathtt{Const}_c &=& \determstateless\big((), \true, c\big) \\[1ex]
  \mathtt{Add} &=& \determstateless\big((x,y), \true, x+y\big) \\[1ex]
  \mathtt{Split} &=& \determstateless\big(x, \true, (x,x)\big) \\[1ex]
  \mathtt{Div} &=& \determstateless\big((x,y),y\not=0,\dfrac{x}{y}\big)
  \end{array}
$$

\subsection{Quantified Linear Temporal Logic Components}
\label{subsec:qltl}

Although powerful, STS components have limitations. In particular, they
cannot express {\em liveness} properties~\cite{AlpernSchneider1985}.
To remedy this, we introduce another type of components, based on 
Linear Temporal Logic (LTL)~\cite{DBLP:conf/focs/Pnueli77} and
quantified propositional LTL (QPTL) \cite{SistlaVardiWolper87,Kesten2002}, which 
extends LTL with $\exists$ and $\forall$ quantifiers over propositional variables.
In this paper we use {\em quantified first-order LTL} (which we abbreviate as QLTL).
QLTL further extends QPTL with functional and relational symbols over 
arbitrary domains, quantification of variables over these domains,
and a next operator applied to variables.\footnote{
A logic similar to the one that we use here is presented in \cite{Kesten1994}, however in \cite{Kesten1994} the next operator can be applied only once to variables, and the logic from \cite{Kesten1994} uses also past temporal operators.
}
We need this expressive power in order to be able to handle general models (e.g.,
Simulink) which often use complex arithmetic formulas,
and also to be able to translate STS
components into semantically equivalent QLTL components (see \S\ref{sec_sts2qltl}).

\subsubsection{QLTL}

QLTL formulas are generated by the following grammar. We assume a set of constants and functional symbols 
($0,1,\ldots$, $\true, \false$, $+,\ldots$), a set of predicate symbols
 ($=,\le,<,\ldots$), and a set of variable names ($x, y, z, \ldots$).

\begin{definition}[Syntax of QLTL]
\label{def:qltl}
	A QLTL formula $\varphi$ is defined by the following grammar:
	\begin{eqnarray*}
		\begin{array}{llll}
		
		term   & ::= & x\ |\ y\ |\ \ldots\ | & \text{(variable names)} \\
		       &  & 0 \ |\ 1\ |\ \ldots\ | \ \true \ | \ \ldots \ | & \text{(constants)} \\
		       &  & term + term \ | \ \ldots\ |   & \text{(functional symbol application)} \\
		            && \Nextvar term & \text{(next applied to a term)}\\
		\varphi & ::= & term = term \ | \ (term \le term) \ | \ \ldots\ \vert & \text{(atomic QLTL formulas)} \\
		          &    & \neg \varphi\ \vert & \text{(negation)}         \\
		           &   & \varphi \lor \psi\ \vert & \text{(disjunction)}  \\
		           &   & \varphi \Until \psi\ \vert & \text{(until)} \\
		           &   & \forall x: \varphi & \text{(forall)}  
		\end{array}
	\end{eqnarray*}
\end{definition}

As in standard first order logic, the {\em bounded variables} of a formula $\varphi$ are the variables in scope of
the universal quantifier $\forall$, and the {\em free variables} of $\varphi$ are  those that are not bounded.
The logic connectives $\land$, $\impl$ 
and $\eqv$ can be expressed with $\neg$ and $\lor$. 
Quantification is over atomic variables.
The existential quantifier $\exists$ can be defined via the universal 
quantifier usually as $\neg \forall \neg$.
The primitive temporal operators are {\em next for terms} ($\Nextvar$) and {\em until} ($\Until$). 
As is standard, QLTL formulas are evaluated over infinite traces, and
$\varphi \Until \psi$ intuitively means that $\varphi$ continuously holds
until some point in the trace where $\psi$ holds.

Formally, we will define the relation $\sigma \models \varphi$
($\sigma$ {\em satisfies} $\varphi$) for a QLTL formula $\varphi$
over free variables $x,y,\ldots$, and an infinite sequence 
$\sigma\in\Sigma^\omega$, where
$\Sigma = \Sigma_x \times \Sigma_y \times \ldots$, and  $\Sigma_x, \Sigma_y,\ldots$ are the types (or {\em domains}) of variables $x,y,\ldots$.
As before we assume that $\sigma$ can be written as a tuple of sequences $(\sigma_x,\sigma_y,\ldots)$
where $\sigma_x\in\Sigma_x^\omega, \sigma_y\in\Sigma_y^\omega,\ldots$. 
The semantics of a term $t$ on variables $x,y,\ldots$ is a function from infinite sequences to infinite sequences 
$\semterm{t}:\Sigma^\omega\to\Sigma_t^\omega$, where $\Sigma = \Sigma_x \times \Sigma_y \times \ldots$, and $\Sigma_t$ is the type of $t$.
When giving the semantics of terms and formulas we assume that constants, functional symbols, and predicate symbols have the standard semantics. For example, we assume that $+,\le,\ldots$ on numeric values have the semantics of standard
arithmetic.

\begin{definition}[Semantics of QLTL]
\label{def:ltl_properties}
	Let $x$ be a variable, $t,t'$ be terms, $\varphi, \psi$ be QLTL formulas, $P$ be a predicate symbol, $f$ be a functional symbol,
	$c$ be a constant, and $\sigma \in \Sigma^{\omega}$ be an infinite sequence.
	$$
	\begin{array}{lll}
		\semterm{x}(\sigma) & := &  \sigma_x \\
		\semterm{c}(\sigma) & := & (\lambda i : c) \\
		\semterm{f(t,t')}(\sigma) & := & (\lambda i: f(\semterm{t}(\sigma)(i), \semterm{t'}(\sigma)(i)))\\
		\semterm{\Nextvar t}(\sigma) & := & \semterm{t}(\sigma^1)\\[1ex]
		\sigma \models P(t,t') & := & P(\semterm{t}(\sigma)(0), \semterm{t'}(\sigma)(0))\\
		\sigma \models \neg \varphi & := & \neg\ \sigma \models \varphi\\
		\sigma \models \varphi \lor \psi & := & \sigma \models \varphi \lor \sigma \models \psi\\
		\sigma \models \varphi \Until \psi & := & (\exists n\ge 0: (\forall\ 0 \leq i < n: \sigma^i \models \varphi) \land \sigma^n \models \psi)\\
		\sigma \models (\forall x: \varphi) & := & (\forall \sigma_x\in\Sigma_x^\omega: (\sigma_x,\sigma) \models \varphi)
	 
	\end{array}
	$$
	\delete{
	\begin{enumerate}
		\item $\semterm{x}(\sigma) \ :=\  \sigma_x$
		\item 
			$\semterm{c}(\sigma) \ := \ (\lambda i : c)$
		\item $\semterm{f(t,t')}(\sigma) \ := \ (\lambda i: f(\semterm{t}(\sigma)(i), \semterm{t'}(\sigma)(i))$
		\item $\semterm{\Nextvar t}(\sigma) \ := \ \semterm{t}(\sigma^1)$
		\item $\sigma \models P(t,t') \ := \ P(\semterm{t}(\sigma)(0), \semterm{t'}(\sigma)(0))$
		\item $\sigma \models \neg \varphi \ := \ \neg\ \sigma \models \varphi$
		\item $\sigma \models \varphi \lor \psi \ := \ \sigma \models \varphi \lor \sigma \models \psi$
		\item $\sigma \models \varphi \Until \psi \ := \ (\exists n: (\forall i < n: \sigma^i \models \varphi) \land \sigma^n \models \psi)$
		\item $\sigma \models (\forall x: \varphi) \ := \ (\forall \sigma_x\in\Sigma_x^\omega: (\sigma_x,\sigma) \models \varphi)$
	\end{enumerate}}
\end{definition}

Other temporal operators can be defined as follows.  {\em Eventually} ($\Evetl \varphi = \true \Until \varphi$) states that $\varphi$ must hold in some future step.  {\em Always} ($\Always \varphi = \neg \Evetl \neg \varphi$) states that $\varphi$ must hold at all steps. 
The next operator for formulas $\Next$ can be defined using the next operator for terms $\Nextvar$.
The formula $\Next \varphi$ is obtained by replacing all occurrences of the free variables in $\varphi$ by their
next versions (i.e., $x$ is replaced by $\Nextvar x$,
$y$ by $\Nextvar y$, etc.).
For example the propositional LTL formula $\Next (x \land \Next y \impl \Always z)$ can be expressed as 
$$(\Nextvar x=\true \land (\Nextvar \Nextvar y = \true) \impl \Always (\Nextvar z=\true)).$$

We additionally introduce the operator: $\varphi\Leads\psi := \neg (\varphi \Until \neg \psi)$. Intuitively, $\varphi\Leads\psi$ holds if whenever $\varphi$ holds continuously up to some step $n-1$, $\psi$ must hold at step $n$.

Two QLTL formulas $\varphi$ and $\psi$ are semantically equivalent, denoted $\varphi \iff \psi$, if  
$$\forall \sigma : (\sigma \models \varphi) \iff (\sigma \models \psi).$$

\begin{lemma}\label{lem:leads}
	Let $\varphi$ be a QLTL formula. Then:
	\begin{enumerate}
		\item $(\exists x: \Always \varphi) \iff \Always (\exists x: \varphi)$ when $\varphi$ does not contain temporal operators.
		\item $\varphi \Leads \varphi \iff \Always \varphi$
		\item $\true \Leads \varphi \iff \Always \varphi$
		\item $\varphi \Leads \true \iff \true $
		\item $\varphi \Leads \false \iff \false $
		\item $\forall y: (\varphi \Leads \psi) \iff (\exists y:\varphi) \Leads \psi$, 
			when $\varphi$ does not contain temporal operators and $y$ is not free in $\psi$.
	\end{enumerate}
\end{lemma}

The proof of the above result, as well as of most results that follow, is omitted. All omitted proofs have been formalized and proved in the Isabelle proof assistant, and are available as part of the public distribution of RCRS from 
\url{http://rcrs.cs.aalto.fi}.
In particular, all results contained in this paper can be accessed from the theory \texttt{RCRS\_Overview.thy} -- either directly in that file or via references to the other RCRS files.

\delete{
\red{should we also say that there is a document available which links the
lemmas etc in this paper to the places in the distribution where these and
their proofs can be found?} \red{Viorel: Here it could be the link to the RCRS\_Overview.html file. We know it after the html output is integrated in the distribution.}
}

\paragraph{Examples.}
Using QLTL we can express {\em safety}, as well as {\em liveness requirements}. Informally, 
a safety requirement expresses that something bad never happens. An example is the formula 
$$\mathtt{thermostat} = \Always (180^\circ \leq t \land t \leq 220^\circ),$$ 
which states that the thermostat-controlled temperature $t$ stays always between $180^\circ$ and $220^\circ$. 

A liveness requirement informally says that something good eventually happens. An example is the formula $\Evetl (t > 200^\circ)$ stating that the temperature $t$ is eventually
over $200^\circ$.

A more complex example is a formula modeling an oven that starts increasing the temperature from an initial value of $20^\circ$ until it reaches $180^\circ$, and then keeps it between $180^\circ$ and $220^\circ$.
$$
\mathtt{oven} = (t = 20^\circ \land ((t < \Nextvar t \land t < 180) \Until \mathtt{thermostat})).
$$
In this example the formula $t < \Nextvar t$ specifies that the temperature increases from some point to the next.

\subsubsection{QLTL Components}

A QLTL component is an atomic component where the input-output behavior is specified by a QLTL formula:

\begin{definition}[QLTL component]
\label{def:qltl_comp}
	A {\em QLTL component} is a tuple $\qltl(x:\Sigma_x, y:\Sigma_y, \varphi)$, 
	where $x,y$ are input and output variables (or tuples of variables)
	of types $\Sigma_x, \Sigma_y$,
        and $\varphi$ is a QLTL formula over $x$ and $y$.
\end{definition}

Intuitively a QLTL component $C=\qltl(x,y,\varphi)$ represents a system that takes as input an infinite sequence $\sigma_x\in\Sigma_x^\omega$ and 
produces as output an infinite sequence $\sigma_y\in\Sigma_y^\omega$ such that $(\sigma_x,\sigma_y) \models \varphi$. If there is no $\sigma_y$ such that 
$(\sigma_x,\sigma_y) \models \varphi$ is true, then input $\sigma_x$ is illegal for $C$, i.e., $C$ is not input-receptive.
There could be many possible $\sigma_y$ for a single $\sigma_x$, in which case
the system is non-deterministic.

As a simple example, we can model the oven as a QLTL component with no input 
variables and the temperature as the only output variable:
$$
\qltl((),t,\mathtt{oven})
$$

\subsection{Well Formed Composite Components}
\label{subsec:well-formed}

Not all composite components generated by the grammar introduced in \S\ref{subsec:component_grammar} are {\em well formed}. Two components $C$ and $C'$ can be composed in series only if the number of outputs of $C$ matches the number of inputs of $C'$, and in addition the input types of $C'$ are the same as the corresponding output types of $C$. 
Also, $\synfb$ can be applied to a component $C$ if the type of the first output of $C$ is the same as the type of its first input.
Formally, for every component $C$ we define below $\intype(C)$ - the 
{\em input type} of $C$, $\outtype(C)$ - the {\em output type} of $C$, 
and $\wf(C)$ - the {\em well-formedness} of $C$, by induction on the structure
of $C$:

$$
\begin{array}{lll}
\intype(\sts(x:\Sigma_x, y:\Sigma_y, s:\Sigma_s, init, trs)) & = & \Sigma_x \\
\intype(\stateless(x:\Sigma_x, y:\Sigma_y, trs)) & = & \Sigma_x \\
\intype(\determ(x:\Sigma_x, s:\Sigma_s, a, inpt, next, out:\Sigma_y)) & = & \Sigma_x \\
\intype(\determstateless(x:\Sigma_x, inpt, out:\Sigma_y)) & = & \Sigma_x \\
\intype(\qltl(x:\Sigma_x, y:\Sigma_y, \varphi)) & = & \Sigma_x \\
\intype(C\; \synsop\; C') & = & \intype(C) \\
\intype(C \synpop C') & = & \intype(C) \times \intype(C') \\
\intype(\synfb (C)) & = & X_2 \times \cdots \times X_n \mbox{ provided } \intype(C) = X_1 \times \cdots \times X_n \\
	& & \text{ for some } \ n \ge 1
\end{array}
$$

$$
\begin{array}{lll}
\outtype(\sts(x:\Sigma_x, y:\Sigma_y, s:\Sigma_s, init, trs)) & = & \Sigma_y \\
\outtype(\stateless(x:\Sigma_x, y:\Sigma_y, trs)) & = & \Sigma_y \\
\outtype(\determ(x:\Sigma_x, s:\Sigma_s, a, inpt, next, out:\Sigma_y)) & = & \Sigma_y \\
\outtype(\determstateless(x:\Sigma_x, inpt, out:\Sigma_y)) & = & \Sigma_y \\
\outtype(\qltl(x:\Sigma_x, y:\Sigma_y, \varphi)) & = & \Sigma_y \\
\outtype(C\; \synsop\; C') & = & \outtype(C') \\
\outtype(C \synpop C') & = & \outtype(C) \times \outtype(C') \\
\outtype(\synfb (C)) & = & Y_2 \times \cdots \times Y_n \mbox{ provided } \outtype(C) = Y_1 \times \cdots \times Y_n \\
	& & \text{ for some } n \ge 1
\end{array}
$$

$$
\begin{array}{lll}
\wf(\sts(x, y, s, init, trs)) & = & true \\
\wf(\stateless(x, y, trs)) & = & true\\
\wf(\determ(x, s, a, inpt, next, out)) & = & true\\
\wf(\determstateless(x, inpt, out)) & = & true\\
\wf(\qltl(x, y, \varphi)) & = & true \\
\wf(C\; \synsop\; C') & = & \wf(C) \land \wf(C') \land \outtype(C) = \intype(C') \\
\wf(C \synpop C') & = & \wf(C) \land \wf(C') \\
\wf(\synfb (C)) & = & \wf(C) \land \intype(C) = X \times X_1 \cdots \times X_n 
	\\ & &
 	\land \; \outtype(C) = X \times Y_1 \cdots \times Y_m,
	\text{ for some } n,m\ge 0.
\end{array}
$$
In the definition above, both $n$ and $m$ are natural numbers.
If $n=0$ then $X_1\times\cdots\times X_n$ denotes the $\Unit$ type.

Note that atomic components are by definition well-formed.
The composite components considered in the sequel are 
required to be well-formed too.

We note that the above well-formedness conditions are not restrictive.
Components that do not have matching inputs and outputs can still be
composed by adding appropriate {\em switching} components which reorder
inputs, duplicate inputs, and so on. An example of such a component
is the component $\mathtt{Split}$, introduced earlier.
As another example, consider the diagram in Fig.~\ref{fig_diagram_switch}:

\begin{figure}[h]
  \centering
  \begin{tikzpicture}
    \node[draw, minimum height=5ex] (A) {$A$};
    \node[draw, minimum height=5ex, right = 6.1ex of A] (B) {$B$};
    \node[draw, minimum height=5ex, right = 6ex of B] (C) {$C$};
  
    \draw[-latex'] ([xshift=-3ex]A.west) -- (A.west);
  
    \draw[-latex'] ([yshift=1ex]A.east) --++ (3ex,0) node[circle,minimum size=3pt,inner sep=0pt,fill=black] {} coordinate(aux1) -- (aux1 -| B.west);
    \draw[-latex'] (aux1) -- ([yshift=-1ex]aux1 |- B.south) coordinate (aux2) -- ([xshift=-2ex]aux2 -| C.west) coordinate (aux3) -- ([yshift=-1ex]aux3 |- C.west) coordinate(aux4) -- (aux4 -| C.west);
    \draw[-latex'] ([yshift=-1ex]A.east) coordinate (aux5) -- (aux5 -| B.west);
    \draw[-latex'] ([yshift=1ex]B.east) coordinate (aux6) -- (aux6 -| C.west);
  
    \draw[-latex'] (C.east) --++ (3ex,0) coordinate(aux7);
    \draw[-latex'] ([yshift=-1ex]B.east) --++ (2ex,0) coordinate(aux8) -- ([yshift=-2ex]aux8 |- C.south) coordinate (aux9) -- (aux9 -| aux7);
  
  \end{tikzpicture}
  \caption{Another block diagram.}
\label{fig_diagram_switch}
\end{figure}
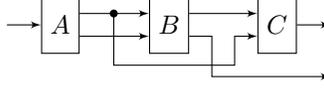

This diagram can be expressed in our language as the composite component:
$$
\begin{array}{l}
A \synsop \mathtt{Switch1} \synsop (B \synpop \id) \synsop \mathtt{Switch2} \synsop (C \synpop \id)
\end{array}
$$
where
\begin{eqnarray*}
\mathtt{Switch1} & = & \determstateless((x,y),\true, (x,y,x)) \\ 
\mathtt{Switch2} & = & \determstateless((u,v,x),\true, (u,x,v))
\end{eqnarray*}

%% file: semantics.tex

\section{Semantics}
\label{sec:semantics}

In RCRS, the semantics of components is defined in terms of
{\em monotonic property transformers} (MPTs). This is inspired
by classical refinement calculus~\cite{backwright:98}, where
the semantics of sequential programs is defined in terms of monotonic
{\em predicate} transformers~\cite{dijkstra:75}.
Predicate transformers are functions that transform sets of {\em post}-states (states reached after the program executes) into sets of {\em pre}-states (states from which the program begins). 
Property transformers map sets of {\em output traces} (that a component
produces) into sets of {\em input traces} (that a component consumes).

In this section we define MPTs formally, and introduce some basic operations 
on them, which are necessary for giving the semantics of components.
The definitions of some of these operations (e.g., product and fusion) are
simple extensions of the corresponding operations on predicate transformers~\cite{backwright:98,BackButler1995}.
Other operations, in particular those related to feedback, are new
(\S\ref{sec_novel_semantical_operators_for_feedback}).
The definition of component semantics is also new (\S\ref{subsec:comp_semantics}).

\subsection{Monotonic Property Transformers}
\label{subsec:mpt}
 
A property transformer is a function $S:(\Sigma_{y}^{\omega} \to \Bool)\to (\Sigma_{x}^{\omega} \to \Bool)$, where $\Sigma_x,\Sigma_y$ are input and output types of the component in question. 
Note that $x$ is the input and $y$ is the output.
A property transformer has a weakest precondition interpretation: it is applied to a set of output traces $Q \subseteq \Sigma_y^\omega$, and 
returns a set of input traces $P \subseteq \Sigma_x^\omega$, such that all traces in $P$ are legal and, when fed to the component, are guaranteed
to produce only traces in $Q$ as output. 

Interpreting properties as sets, monotonicity of property transformers
simply means that these functions are monotonic with respect to set inclusion.
That is, $S$ is {\em monotonic} if for any two properties $q,q'$, if $q \subseteq q'$ then $S(q) \subseteq S(q')$.

For an MPT $S$ we define its set of {\em legal input traces} as $\legal(S) = S(\top)$, where $\top$ is the greatest predicate extended to traces.
Note that, because of monotonicity, and the fact that $q\subseteq\top$ holds
for any property $q$, we have that $S(q)\subseteq\legal(S)$ for all $q$.
This justifies the definition of $\legal(S)$ as a ``maximal'' set of 
input traces for which a system does not fail, assuming no restrictions
on the post-condition.
An MPT $S$ is said to be {\em input-receptive} if $\legal(S) = \top$.

\subsubsection{Some Commonly Used MPTs}
\label{subsubsec:mpt_basic}

\begin{definition}[Skip]
  $\Skip$ is defined to be the MPT such that for all $q$, $\Skip(q) = q$.
\end{definition}
$\Skip$ models the identity function, i.e., the system that accepts all
input traces and simply transfers them unchanged to the output
(this will become more clear when we express $\Skip$ in terms of assert
or update transformers, below).
Note that $\Skip$ is different from $\id$, defined above, although the
two are strongly related: $\id$ is a component, i.e., a syntactic object, while 
$\Skip$ is an MPT, i.e., a semantic object.
As we shall see in \S\ref{subsec:comp_semantics},
the semantics of $\id$ is defined as $\Skip$.

\begin{definition}[Fail]
\label{def:fail}
	$\Fail$ is defined to be the MPT such that for all $q$,
	$\Fail(q) = \bot$.
\end{definition}
Recall that $\bot$ is the predicate that returns $\mathsf{false}$ for any
input. Thus, viewed as a set, $\bot$ is the empty set.
Consequently, $\Fail$ can be seen to model a system which rejects all inputs,
i.e., a system such that for any output property $q$, there are
no input traces that can produce an output trace in $q$.

\begin{definition}[Assert]
\label{def:assert}
	Let $p \in \Sigma^{\omega} \to \Bool$ be
	a property. The {\em assert property transformer} $\{p\} : (\Sigma^{\omega} \to \Bool) \to(\Sigma^{\omega} \to \Bool)$ is 
	defined by $$\{p\}(q) = p\land q.$$
\end{definition}
The assert transformer $\{p\}$ can be seen as modeling a system which accepts
all input traces that satisfy $p$, and rejects all others. For all the traces
that it accepts, the system simply transfers them, i.e., it behaves as the
identity function.

To express MPTs such as assert transformers syntactically, let us introduce
some notation. First, we can use lambda notation for predicates, as in
$\lambda (\sigma, \sigma'): (\sigma = \sigma')$ for some predicate $p:\Sigma^\omega\to\Sigma^\omega\to\Bool$
which returns $\mathsf{true}$ whenever it receives two equal traces.
Then, instead of writing $\{\lambda (\sigma, \sigma'): (\sigma = \sigma')\}$ for the corresponding
assert transformer $\{p\}$, we will use the slightly lighter notation
$\{\sigma, \sigma' \mid \sigma = \sigma'\}$.

\begin{definition}[Demonic update]
\label{def:demonic}
	Let $r:\Sigma_x^{\omega} \to \Sigma_y^{\omega} \to \Bool$ be a relation. The {\em demonic update property transformer} 
	$[r]: (\Sigma_{y}^{\omega} \to \Bool) \to (\Sigma_{x}^{\omega} \to \Bool)$ is defined by 
	$$[r](q) = \{\sigma\ \vert\ \forall \sigma': r(\sigma)(\sigma') \impl \sigma' \in q \}.$$
\end{definition}

That is, $[r](q)$ contains all input traces $\sigma$ which are guaranteed
to result into an output trace in $q$ when fed into the (generally
non-deterministic) input-output relation $r$.
The term ``demonic update'' comes from the refinement calculus literature~\cite{backwright:98}.

Similarly to assert, we introduce a lightweight notation for the demonic update.
If $r$ is an expression in $\sigma$ and $\sigma'$, then 
$[\sigma \leadsto \sigma' \mid r] = [\lambda (\sigma, \sigma'): r]$. 
For example, 
$[\sigma_x, \sigma_y \leadsto \sigma_z\ \vert\ \forall i: \sigma_z(i) = \sigma_x(i) + \sigma_y(i)]$
is the system which produces as output the sequence $\sigma_z= (\lambda i: \sigma_x(i)+\sigma_y(i))$, 
where $\sigma_x$ and $\sigma_y$ are the input sequences.
If $e$ is an expression in $\sigma$, then $[\sigma \leadsto e]$ is
defined to be $[\sigma \leadsto \sigma' \ | \ \sigma' = e]$, where $\sigma'$ is a new variable different from $\sigma$ and which does not occur free in $e$.
For example, $[\sigma \leadsto (\lambda i :\sigma(i) + 1)] = [\sigma \leadsto \sigma'\ \vert\ \sigma' = (\lambda i :\sigma(i) + 1)]$.

The following lemma states that $\Skip$ can be defined as an assert transformer,
or as a demonic update transformer.

\begin{lemma}
  $\Skip = [\sigma \leadsto \sigma' \ \vert\  \sigma = \sigma']= \{\top\} = \{\sigma \mid \mathsf{true}\}$.
\end{lemma}

In general $\Skip$, $\Fail$, and other property transformers are polymorphic with respect to their input and output types.  
In $\Skip$ the input and output types must be the same.
$\Fail$, on the other hand, may have an input type and a different output type.

\begin{definition}[Angelic update]
\label{def:angelic}
  Let $r:\Sigma_x^{\omega} \to \Sigma_y^{\omega} \to \Bool$ be a relation. The {\em angelic update property transformer} 
  $\{r\}: (\Sigma_{y}^{\omega} \to \Bool) \to (\Sigma_{x}^{\omega} \to \Bool)$ is defined by
  $$\{r\}(q) = \{ \sigma\ \vert\ \exists \sigma': r(\sigma)(\sigma') \land \sigma' \in q\}.$$
\end{definition}

An input sequence $\sigma$ is in $\{r\}(q)$ if there exists an output sequence $\sigma'$ such that $r(\sigma)(\sigma')$ and $\sigma' \in q$.
Notice the duality between the angelic and demonic update transformers.
Consider, for example, a relation $r = \{(\sigma, \sigma'), (\sigma, \sigma'')\}$. If $q = \{\sigma', \sigma'' \}$, then $\{r\}(q) = [r](q) = \{\sigma\}$.
If $q =\{\sigma'\}$ then $\{r\}(q) = \{\sigma\}$, while $[r](q) = \emptyset$.

We use a lightweight notation for the angelic update transformer, similar to the one for demonic update. If $r$ is an expression in $\sigma$ and $\sigma'$, 
then $\{\sigma \leadsto \sigma'\ \vert\ r \} = \{\lambda (\sigma, \sigma'): r \}$.

\begin{lemma}
Assert is a particular case of angelic update: $\{p\} = \{\sigma \leadsto \sigma' \ | \ 
p(\sigma) \land \sigma = \sigma'\}$. 
\end{lemma}

\subsubsection{Operators on MPTs: Function Composition, Product, and Fusion}

As we shall see in \S\ref{subsec:comp_semantics}, the semantics of composition
operators in the language of components will be defined by the corresponding
composition operators on MPTs.
We now introduce the latter operators on MPTs.
First, we begin by the operators that have been known in the literature,
and are recalled here. In \S\ref{sec_novel_semantical_operators_for_feedback} we introduce some novel operators explicitly
designed in order to handle feedback composition.

Serial composition of MPTs (and property transformers in general) is simply
function composition. 
Let $S: (\Sigma^{\omega}_y \to \Bool) \to (\Sigma^{\omega}_x \to \Bool)$ and $T: (\Sigma^{\omega}_z \to \Bool) \to (\Sigma^{\omega}_y \to \Bool)$ be two property transformers. 
Then $S \semsop T : (\Sigma^{\omega}_z \to \Bool) \to (\Sigma^{\omega}_x \to \Bool)$, 
is the function composition of $S$ and $T$, i.e., $\forall q: (S \semsop T) (q) = S(T(q))$.
Note that serial composition preserves monotonicity, so that if $S$ and $T$
are MPTs, then $S \semsop T$ is also an MPT.
Also note that $\Skip$ is the neutral element for serial composition, i.e., $S \semsop \Skip = \Skip \semsop S = S$. 

To express parallel composition of components, we need a kind of Cartesian
product operation on property transformers. We define such an operation
below. Similar operations for predicate transformers have been
proposed in~\cite{BackButler1995}.

\begin{definition}[Product]
\label{def:sempop}
	Let $S: (\Sigma^{\omega}_y \to \Bool) \to (\Sigma^{\omega}_x \to \Bool)$ and 
	$T: (\Sigma^{\omega}_v \to \Bool) \to (\Sigma^{\omega}_u \to \Bool)$.
	The {\em product} of $S$ and $T$, denoted $S \sempop T : (\Sigma^{\omega}_y
	\times \Sigma^{\omega}_v \to \Bool)  \to (\Sigma^{\omega}_x\times \Sigma^{\omega}_u \to \Bool)$, 
	is given by 
	$$(S \sempop T)(q) = \{(\sigma, \sigma')\ \vert\ \exists p: \Sigma^{\omega}_y \to \Bool, p' : \Sigma^{\omega}_v \to \Bool : p \times p' \subseteq q \land \sigma \in S(p) \land \sigma' \in T(p') \}$$
	where $(p\times p') (\sigma_y,\sigma_v) = p(\sigma_y)\land p'(\sigma_v)$.
\end{definition}

\begin{lemma}
 For arbitrary $S$ and $T$, $S \sempop T$ is monotonic. 
\end{lemma}

The neutral element for the product composition is the $\Skip$ MPT that has $\Unit$ as input and output type. 

In order to define a feedback operation on MPTs, we first define two 
auxiliary operations: {\em Fusion} and {\em IterateOmega}.
Fusion is an extension of a similar operator introduced 
previously for predicate transformers in~\cite{BackButler1995}.
IterateOmega is a novel operator introduced in the sequel.

\begin{definition}[Fusion]
\label{def:fusion}
	If $S = \{S_i\}_{i \in I}$, $S_i : (\Sigma^{\omega}_y \to \Bool) \to (\Sigma^{\omega}_x \to \Bool)$ is a collection of MPTs, then the {\em fusion} of $S$ is the MPT 
	$\Fusion_{i \in I} (S_i) : (\Sigma^{\omega}_y \to \Bool) \to (\Sigma^{\omega}_x \to \Bool)$ defined by
	$$
	(\Fusion_{i \in I} (S_i)) (q) = \{\sigma\ \vert\ \exists p:I\to\Sigma_y^{\omega}\to\Bool : \bigcap_i p_i  \subseteq q \land \sigma \in \bigcap_i S_i(p_i)\} 
	$$
\end{definition}

The $\Fusion$ operator satisfies the following property.
\begin{lemma} 
  For $I\not = \emptyset$ we have $$\Fusion_{i\in I} (\{p_i\}\semsop [r_i]) = \{\bigcap_{i\in I} p_i\} \semsop [\bigcap_{i\in I} r_i].$$
\end{lemma}

\subsubsection{Novel Operators Used in Semantical Definition of Feedback}
\label{sec_novel_semantical_operators_for_feedback}

The \textsf{IterateOmega} operator\delete{, illustrated in Fig.~\ref{fig:itom},} is defined as follows:

\begin{definition}[IterateOmega]
\label{def:itom}
	$$\IterateOmega (S) = \Fusion_{n \in \Nat} (S^n \circ [\sigma \leadsto \sigma'\ \vert\ \forall i: i + 1 < n \impl \sigma_i = \sigma'_i])$$
\end{definition}

The {\em feedback} operator consists of connecting the first output of an MPT $S$ with its first input. Formally, feedback is defined as follows.

\begin{definition}[Feedback]
	Let $S : (\Sigma_u\times\Sigma_y^\omega \to \Bool) \to (\Sigma_u\times\Sigma_x^\omega \to \Bool)$ be an MPT. The {\em feedback operator} on $S$, denoted $\semfb(S)$, is given by the MPT  
	$$
	\begin{array}{lll} 
	\semfb(S) & = & 
		\{\sigma_x \leadsto \sigma_u, \sigma_y, \sigma_x\} \\ && \semsop \  \IterateOmega 
		 ([\sigma_u, \sigma_y, \sigma_x \leadsto \sigma_u, \sigma_x, \sigma_x] \semsop  
		 (S \sempop \Skip)) \\ 
		 && \semsop \ [\sigma_u, \sigma_y, \sigma_x \leadsto \sigma_y ] 
	\end{array}
	$$	
\end{definition}

\paragraph{Example.}
As an example we show how to derive $\semfb(S)$ for
$S = [\sigma_u,\sigma_x\leadsto 0 \cdot \sigma_x + 0 \cdot\sigma_u,0 \cdot \sigma_x + 0 \cdot\sigma_u]$, where
$\sigma + \sigma' = (\lambda i:\sigma(i) + \sigma'(i))$, and $0\cdot\sigma$ is 0 concatenated with $\sigma$. 
For now, we note that $S$ is the semantics of the composite component
$\mathtt{Add} \synsop \mathtt{UnitDelay} \synsop \mathtt{Split}$,
which corresponds to the inner part of the diagram of Fig.~\ref{fig:diagram},
before applying feedback. We will complete the formal definition of the
the semantics of this diagram in \S\ref{subsec:comp_semantics}. For now,
we focus on deriving $\semfb(S)$, in order to illustrate how the 
$\semfb$ operator works.

Let
$$T = [\sigma_u, \sigma_y, \sigma_x \leadsto \sigma_u, \sigma_x, \sigma_x] \semsop  
		 (S \sempop \Skip ) = 
[\sigma_u,\sigma_y, \sigma_x \leadsto 0 \cdot \sigma_x + 0 \cdot\sigma_u, 0 \cdot \sigma_x + 0 \cdot\sigma_u, \sigma_x]
$$
Then, we have
$$
\begin{array}{lll}
T \circ T & = & [\sigma_u,\sigma_y, \sigma_x \leadsto 
0 \cdot \sigma_x + 0\cdot 0 \cdot \sigma_x + 0\cdot 0 \cdot\sigma_u, 
0 \cdot \sigma_x + 0\cdot 0 \cdot \sigma_x + 0\cdot 0 \cdot\sigma_u, \sigma_x]\\
\ldots \\
T^n & = & [\sigma_u,\sigma_y, \sigma_x \leadsto 
0 \cdot \sigma_x + \ldots + 0^n \cdot \sigma_x + 0^n \cdot\sigma_u, 
0 \cdot \sigma_x + \ldots + 0^n \cdot \sigma_x + 0^n \cdot\sigma_u, \sigma_x]
\end{array}
$$
where $0^n$ is a finite sequence of $n$ $0$s.
 We also have
$$
\begin{array}{ll}
& T^n \circ [\sigma \leadsto \sigma'\ \vert\ \forall i: i + 1 < n \impl \sigma_i = \sigma'_i] \\
= \\
&[\sigma_u,\sigma_y, \sigma_x \leadsto \sigma \ |\ 
\forall i: i+1<n \impl
\sigma(i) = (\Sigma_{j<i}\sigma_x(j),\Sigma_{j<i}\sigma_x(j), \sigma_x(i))]
\end{array} 
$$
\delete{
\blue{Note that $\sigma$ in the expression above is explicitly given for $i +1 < n$. 
$\sigma_u$ is preceded by $0^n$, and therefore the value $0$ can be dropped from the sum for index $i$.
For $i +1 \geq n$, $\sigma^i$ holds the values computed by $T^n$.}
}
Then
$$
\begin{array}{ll}
	& \semfb(S) \\ 
	= \\
& \{\sigma_x \leadsto \sigma_u, \sigma_y, \sigma_x\}  
		\semsop \IterateOmega(T) 
		\semsop \ [\sigma_u, \sigma_y, \sigma_x \leadsto \sigma_y ] \\

= \\
& \{\sigma_x \leadsto \sigma_u, \sigma_y, \sigma_x\}  
		\semsop \Fusion_{n\in\Nat}(T^n \circ [\sigma \leadsto \sigma'\ \vert\ \forall i: i + 1 < n \impl \sigma_i = \sigma'_i]) 
		\semsop \ [\sigma_u, \sigma_y, \sigma_x \leadsto \sigma_y ] \\

= \\
	&	\{\sigma_x \leadsto \sigma_u, \sigma_y, \sigma_x\}  \semsop \  
[\sigma_u,\sigma_y, \sigma_x \leadsto \sigma \ |\ 
\forall i: 
\sigma(i) = (\Sigma_{j<i}\sigma_x(j),\Sigma_{j<i}\sigma_x(j), \sigma_x(i))]		
		  \semsop \ [\sigma_u, \sigma_y, \sigma_x \leadsto \sigma_y ] \\
  = \\
  & 
  [\sigma_x \leadsto \sigma_y \ |\ 
\forall i: \sigma_y(i) = \Sigma_{j<i}\sigma_x(j)]
\end{array} 
$$
Finally we obtain
\begin{equation}\label{eq:ex-feedback}
 \semfb(S) =
  [\sigma_x \leadsto \sigma_y \ |\ \forall i: \sigma_y(i) = \Sigma_{j<i}\sigma_x(j)]
\end{equation}
This is the system that outputs the trace $(\lambda i: \Sigma_{j<i}\sigma_x(j))$ for input trace $\sigma_x$.

\subsubsection{Refinement}

A key element of RCRS, as of other compositional frameworks, is the notion
of refinement, which enables substitutability and other important concepts
of compositionality. Semantically, refinement is defined as follows:

\begin{definition}[Refinement]
	Let $S,T : (\Sigma_y^\omega \to \Bool) \to (\Sigma_x^\omega \to \Bool)$ be two MPTs. 
	We say that $T$ {\em refines} $S$ (or that $S$ {\em is refined by} $T$), written $S \semref T$, if and only if $\forall q: S(q) \subseteq T(q)$.
\end{definition}

All operations introduced on MPTs preserve the refinement relation:
\begin{theorem}
 If $S,T,S',T'$ are MPTs of appropriate types such that $S\sqsubseteq S'$ and $T \sqsubseteq T'$, then
 \begin{enumerate}
  \item $S\semsop T \sqsubseteq S'\semsop T' $
  and $S\sempop T \sqsubseteq S'\sempop T' $
  and $\Fusion(S, T) \sqsubseteq \Fusion(S', T') $ (\cite{BackButler1995,backwright:98})
  \item $\IterateOmega(S) \sqsubseteq \IterateOmega(S') $
  \item $\semfb(S) \sqsubseteq \semfb(S') $
 \end{enumerate}
\end{theorem}

\subsection{Subclasses of MPTs}
\label{subsec:mpt_classes}

Monotonic property transformers are a very rich and powerful class of semantic objects. 
In practice, the systems that we deal with often fall into restricted subclasses of MPTs, which are easier to represent syntactically and manipulate symbolically.
We introduce these subclasses next.

\subsubsection{Relational MPTs}
\label{subsubsec:relational_pt}

\begin{definition}[Relational property transformers]
\label{def:rpt}
	A {\em relational property transformer} (RPT) $S$ is an MPT of the form $\{p\} \semsop [r]$. We call $p$ the {\em precondition} of $S$ and $r$ the {\em input-output relation} of $S$.
\end{definition}
Relational property transformers correspond to {\em conjunctive}
transformers \cite{backwright:98}.
A transformer $S$ is conjunctive if it satisfies
$S(\bigcap_{i\in I}q_i) = \bigcap_{i\in I} S(q_i)$ for all 
$(q_i)_{i\in I}$ and $I\not = \emptyset$.

$\Fail$, $\Skip$, any assert transformer $\{p\}$, and any demonic update transformer $[r]$, are RPTs.
Indeed, $\Fail$ can be written as $\{\sigma \ |\ \false\} \semsop [\sigma \leadsto \sigma' \ |\ \true]$.
$\Skip$ can be written as $\{\sigma \ |\ \true\} \semsop [\sigma \leadsto \sigma]$.
The assert transformer $\{p\}$  can be written as the RPT $\{p\} \semsop [\sigma \leadsto \sigma]$.
Finally, the demonic update transformer $[r]$ can be written as the RPT $\{\sigma \ |\ \true\} \semsop [r]$. 
Angelic update transformers are generally not RPTs:
the angelic update transformer $\{\sigma \leadsto \sigma'\ |\ \true\}$ is not an RPT, as it is not conjunctive.

\paragraph{Examples.} 
Suppose we wish to specify a system that performs division. Here
 are two possible ways to represent this system with RPTs:
$$S_1 = \{\top\} \semsop [\sigma_x, \sigma_y \leadsto \sigma_z\ \vert\ \forall i: \sigma_y(i) \not= 0 \land \sigma_z(i) = \frac{\sigma_x(i)}{\sigma_y(i)}]$$
$$S_2 = \{\sigma_x, \sigma_y\ \vert\ \forall i: \sigma_y(i) \not= 0 \} \semsop [\sigma_x, \sigma_y \leadsto \sigma_z\ \vert\ \sigma_z(i) = \frac{\sigma_x(i)}{\sigma_y(i)}]$$
Although $S_1$ and $S_2$ are both relational, they are not equivalent transformers.
$S_1$ is input-receptive: it accepts all input traces. However, if at some
step $i$ the input $\sigma_y(i)$ is $0$, then the output $\sigma_z(i)$
is arbitrary (non-deterministic).
In contrast, $S_2$ is non-input-receptive as it accepts only those traces $\sigma_y$ that are guaranteed to be non-zero at every step, i.e., those that
satisfy the condition $\forall i: \sigma_y(i) \not= 0$.

\begin{theorem}[RPTs are closed under serial, product and fusion compositions]
\label{thm:serial_product_rpt}
	Let $S = \{p\} \semsop [r]$ and $S' = \{p'\} \semsop [r']$ be two RPTs, with $p$, $p'$, $r$ and $r'$ of appropriate types. Then
	$$S \semsop S' = \{\sigma \ | \ p(\sigma) \land (\forall \sigma': r(\sigma)(\sigma') \impl p'(\sigma'))\} \semsop [r \circ r']$$
and
	$$S \sempop S' = \{\sigma_x, \sigma_y \ \vert\  p(\sigma_x) \land p'(\sigma_y)\} \semsop [\sigma_x, \sigma_y 
	 \leadsto \sigma_x', \sigma_y' \ \vert \ r(\sigma_x)(\sigma_x') \land r'(\sigma_y)(\sigma_y')]$$
and 
	 $$\Fusion(S, S') = \{p \land p'\} \semsop [r \land r']$$
\end{theorem}

RPTs are not closed under $\semfb$. For example, we have 
$$
\semfb([\sigma_x,\sigma_z \leadsto \sigma_x, \sigma_x]) = \{\sigma \leadsto \sigma' \ | \ \true\}
$$
which is a non-relational angelic update transformer as we said above.

Next theorem shows that the refinement of RPTs can be reduced to proving a first order property. 

\begin{theorem} \label{refinement-rpt}
  For $p,p',r,r'$ of appropriate types we have.
 $$\{p\}\semsop [r] \sqsubseteq \{p'\}\semsop [r']  \iff (p\subseteq p' \land (\forall \sigma_x : p(\sigma_x) \impl r'(\sigma_x) \subseteq r(\sigma_x)))$$
\end{theorem}

\subsubsection{Guarded MPTs}
\label{subsec:guarded_pt}

Relational property transformers correspond to systems that have natural
syntactic representations, as the composition $\{p\} \semsop [r]$, where
the predicate $p$ and the relation $r$ can be represented syntactically
in some logic.
Unfortunately, RPTs are still too powerful. In particular, they
allow system semantics that cannot be implemented. For example, consider
the RPT $\Magic = [\sigma\leadsto \sigma' \ | \ \mathsf{false}]$.
It can be shown that for any output property $q$ (including $\bot$),
we have $\Magic(q)=\top$. Recall that, viewed as a set, $\top$ is the set
of all traces. This means that, no matter what the post-condition $q$ is,
$\Magic$ somehow manages to produce output traces satisfying $q$ no matter what the input trace is (hence the name ``magic'').
In general, an MPT $S$ is said to be {\em non-miraculous} (or to
satisfy the {\em law of excluded miracle}) if 
$S(\bot) = \bot$.
We note that in \cite{dijkstra:75}, sequential programs are modeled using predicate transformers that are {\em conjunctive} and satisfy the {\em low of excluded miracle}.

We want to further restrict RPTs so that miraculous behaviour does not arise.
Specifically, for an RPT $S = \{p\}\semsop [r]$ and an input sequence $\sigma$, if there is no $\sigma'$ such that $r(\sigma)(\sigma')$ is satisfied, then we want $\sigma$ to be illegal, i.e., we want $p(\sigma) = \mathsf{false}$. 
We can achieve this by taking $p$ to be $\In{r}$.
Recall that if $r: X \to Y \to \Bool$, then
$\In{r}(x) = (\exists y: r(x)(y))$.
Taking $p$ to be $\In{r}$ effectively means that $p$ and $r$ are combined into a single specification $r$ which can also restrict the inputs. This is also the approach followed in the theory of relational interfaces~\cite{TripakisLHL11}.

\begin{definition}[Guarded property transformers]
\label{def:gpt}
	The {\em guarded property transformer} (GPT) of a relation $r$ is an RPT, denoted $\{r]$, defined by $\{r] = \{\In{r}\} \semsop [r]$.
\end{definition}

It can be shown that an MPT $S$ is a GPT if and only if $S$ is conjunctive and non-miraculous \cite{backwright:98}.
$\Fail$, $\Skip$, and any assert property transformer are GPTs.
Indeed, $\Fail = \{\bot]$ and $\Skip = \{\sigma \leadsto \sigma\ |\ \top]$. 
The assert transformer can be written as $\{p\}=\{\sigma \leadsto \sigma\ |\ p(\sigma)]$.
The angelic and demonic update property transformers are generally not GPTs.
The angelic update property transformer is not always conjunctive in order to be a GPT.
The demonic update property transformer is not in general a GPT because is not always non-miraculous ($\Magic(\bot) = \top \not = \bot$).
The demonic update transformer $[r]$ is a GPT if and only if $\In{r} = \top$ and in this case we have $[r]=\{r]$.

\begin{theorem}[GPTs are closed under serial and parallel compositions]
\label{th:serial_product_gpt}
  Let $S = \{r]$ and $S' = \{r']$ be two GPTs with $r$ and $r'$ of appropriate types. Then
  $$ S \semsop S' = \{\sigma_x \leadsto \sigma_z\ |\ \In{r}(\sigma_x) \land \big(\forall \sigma_y: r(\sigma_x)(\sigma_y) \impl \In{r'}(\sigma_y)\big) \land (r \circ r')(\sigma_x, \sigma_z)]$$
  and
  $$ S \sempop S' = \{\sigma_x, \sigma_y \leadsto \sigma'_x, \sigma'_y\ |\ r(\sigma_x)(\sigma'_x) \land r'(\sigma_y)(\sigma'_y)]$$
\end{theorem}

GPTs are not closed under $\Fusion$ neither $\semfb$.
Indeed, we have already seen in the previous section that $\semfb$ applied to the GPT $[\sigma_x,\sigma_z \leadsto \sigma_x,\sigma_x]$ is not an RPT, and therefore not a GPT either.  For the fusion operator, we have $\Fusion([x\leadsto 0],[x\leadsto 1]) = [\bot]$, which is not a GPT.

A corollary of Theorem~\ref{refinement-rpt} is that refinement of GPTs can
be checked as follows:
\begin{corollary}
$$
\{r] \sqsubseteq \{r']  \iff (\In{r}\subseteq \In{r'} \land (\forall \sigma_x : \In{r}(\sigma_x) \impl r'(\sigma_x) \subseteq r(\sigma_x)))
$$
\end{corollary}

\subsubsection{Other subclasses and overview}
\label{subsubsec:subclasses_gpt}

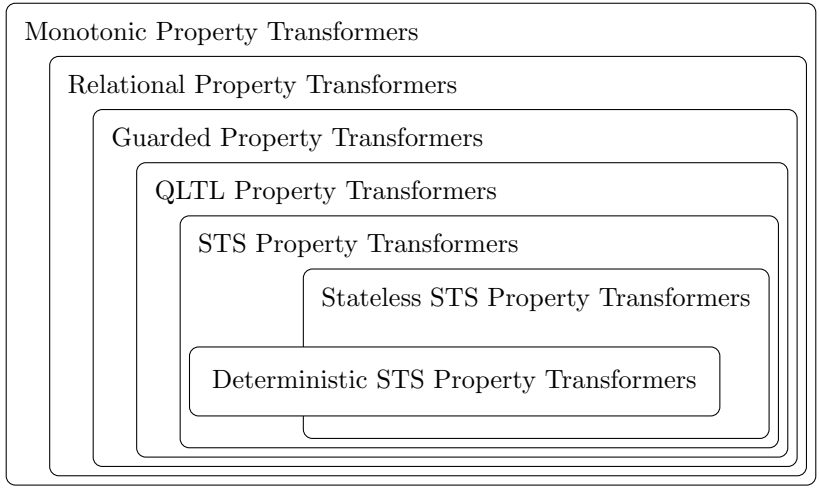
\begin{figure}[h]
	\centering
	\begin{tikzpicture}
	    \node(A){Monotonic Property Transformers};
	    \node[below = 2ex of A.south west](B){};
	    \node[right = 3ex of B](C){Relational Property Transformers};
	    \node[below = 2ex of C.south west](D){};
	    \node[right = 3ex of D](E){Guarded Property Transformers};
	    \node[below = 2ex of E.south west](F){};
	    \node[right = 3ex of F](G){QLTL Property Transformers};

	    \node[below = 2ex of G.south west](H){};
	    \node[right = 3ex of H](I){STS Property Transformers};
	    
	    \node[below = 2ex of I.south west](J){};
	    \node[right = 10ex of J](K){Stateless STS Property Transformers};

	    \node[below = 8ex of K](L){};
	    
	    \node[draw, rounded corners=0.8ex, fit = (K) (L)](M){};

	    \node[draw, rounded corners=0.8ex, below = 7ex of I.south west, anchor = north west, inner sep = 2ex,fill=white](X){Deterministic STS Property Transformers};

	    \node[draw, rounded corners=0.8ex, fit = (I) (X) (M)](N){};

	    \node[draw, rounded corners=0.8ex, fit = (G) (N)](O){};
	    \node[draw, rounded corners=0.8ex, fit = (E) (O)](P){};
	    \node[draw, rounded corners=0.8ex, fit = (C) (P)](Q){};
	    \node[draw, rounded corners=0.8ex, fit = (A) (Q)](R){};
	    
	\end{tikzpicture}
	\caption{Overview of the property transformer classes and their containment relations.}\label{fig:classification}
\end{figure}

The containment relationships among the various subclasses of MPTs are illustrated in
Fig.~\ref{fig:classification}. In addition to the subclasses discussed above,
we introduce several more subclasses of MPTs in the sections that follow,
when we assign semantics (in terms of MPTs) to the various atomic components
in our component language. For instance, QLTL components give rise to
{\em QLTL property transformers}. Similarly,
STS components, stateless STS components, etc., 
give rise to corresponding subclasses of MPTs.
The containment relationships between these classes will be proven in the
sections that follow. For ease of reference, we provide some forward links
to these results also here. The fact that QLTL property transformers are GPTs follows by definition of the semantics of QLTL components: see
\S\ref{subsec:comp_semantics}, equation~(\ref{eq_sem_qltl}).
The fact that STS property transformers are a special case of QLTL property transformers follows from the transformation of an STS component into a semantically equivalent QLTL component: see \S\ref{sec_sts2qltl} and Theorem~\ref{thm_sts2qltl}.
The inclusions for subclasses of STS property transformers follow by definition
of the corresponding components (see also Fig.~\ref{fig:basic}).

\subsection{Semantics of Components as MPTs}
\label{subsec:comp_semantics}

We are now ready to define the semantics of our language of components in
terms of MPTs. 
Let $C$ be a well formed component.
The semantics of $C$, denoted $\sem{C}$, is a property transformer 
of the form:
$$
\sem{C}: ((\outtype(C))^\omega \to \Bool) \to ((\intype(C))^\omega \to \Bool).
$$

We define $\sem{C}$ by induction on the structure of $C$. 
First we give the semantics of QLTL components and composite components:
\begin{eqnarray}
\label{eq_sem_qltl}
 \sem{\qltl(x,y,\varphi)} & = & \{\sigma_x \leadsto \sigma_y \ | \ (\sigma_x,\sigma_y) \models \varphi]\\
 \sem{C \synsop C'} & = & \sem{C} \semsop \sem{C'} \\
 \sem{C \synpop C'} & = & \sem{C} \sempop \sem{C'} \\
 \sem{\synfb(C)} & = & \semfb(\sem{C})  
\end{eqnarray}
The semantics of QLTL components satisfies the following property:
\begin{lemma} \label{lem:sem:qltl}
If $\varphi$ is a LTL formula on variables $x$ and $y$, we have:
$$
\sem{\qltl(x,y,(\exists y:\varphi)\land \varphi)} = 
\sem{\qltl(x,y, \varphi)}
$$
\end{lemma} 

To define the semantics of STS components, we first introduce some
auxiliary notation.

Consider an STS component $C = \sts(x, y, s, \mathit{init\_exp}, \mathit{trs\_exp})$. We define the predicate
$\run_C :\Sigma_s^\omega\times\Sigma_x^\omega\times\Sigma_y^\omega \to \Bool$ as 
$$\run_C(\sigma_s, \sigma_x,\sigma_y) = (\forall i: \mathit{trs\_exp} (\sigma_s(i), \sigma_x(i))(\sigma_s(i+1), \sigma_y(i))).$$
Intuitively, 
if $\sigma_x\in\Sigma_x^\omega$ is the input sequence, $\sigma_y\in\Sigma_y^\omega$ is the output sequence, and $\sigma_s\in\Sigma_s^\omega$ is the sequence of values of state variables, then
$\run_C(\sigma_s, \sigma_x,\sigma_y)$ holds if at each step of the execution,
the current state, current input, next state, and current output, satisfy the
$\mathit{trs\_exp}$ predicate.

We also formalize the illegal input traces of STS component $C$ as follows:
\begin{gather*}
  \illegal_C(\sigma_x) = (\exists \sigma_s, \sigma_y, k: \mathit{init\_exp}(\sigma_s(0))\ \land 
  (\forall i<k: \mathit{trs\_exp} (\sigma_s(i), \sigma_x(i))(\sigma_s(i+1), \sigma_y(i)))\ \land \\ \neg \In{\mathit{trs\_exp}}(\sigma_s(k), \sigma_x(k)))
\end{gather*}

Essentially, $\illegal_C(\sigma_x)$ states that there exists some point in the
execution where the current state and current input violate the precondition
$\In{\mathit{trs\_exp}}$ of predicate $\mathit{trs\_exp}$, i.e., there exist
no output and next state to satisfy $\mathit{trs\_exp}$ for that given
current state and input.

Then, the semantics of an STS component $C$ is given by:
\begin{equation}
\label{eq_sem_sts}
\sem{C} = \{\neg \illegal_C\} \semsop [\sigma_x \leadsto \sigma_y\ |\  
  (\exists \sigma_s : \mathit{init\_exp}(\sigma_s(0)) \land \run_C(\sigma_s, \sigma_x,\sigma_y)) ]
\end{equation}

We give semantics to stateless and/or deterministic STS components using the
corresponding mappings from general STS components.
If $C$ is a stateless STS, $C'$ is a deterministic STS, and $C''$ is a stateless deterministic STS, then:

\begin{eqnarray}
\sem{C} & = & \sem{\stateless 2 \sts(C)} \\
\sem{C'} & = & \sem{\determ 2 \sts(C')} \\
\label{eq_semantics_det_stateless}
\sem{C''} & = & \sem{\determstateless 2\determ (C'')} \delete{\\
\blue{\sem{C''}} & = & \blue{\sem{\determstateless 2\stateless(C'')}}}
\end{eqnarray}

Note that the semantics of a stateless deterministic STS component $C''$ is
defined by converting $C''$ into a deterministic STS component, by
Equation~(\ref{eq_semantics_det_stateless}) above.
Alternatively, we could have defined the semantics of $C''$ by converting it into a stateless STS component, using the mapping  $\determstateless 2\stateless$.
In order for our semantics to be well-defined, we need to show that regardless of which conversion we choose, we obtain the same result.
Indeed, this is shown by the lemma that follows:
\begin{lemma} For a stateless deterministic STS $C''$ we have:
\begin{equation}
\sem{\determstateless 2\determ (C'')} = \sem{\determstateless2\stateless(C'')}
\end{equation}
\end{lemma}

Observe that, by definition, the semantics of QLTL components are GPTs. 
The semantics of STS components are defined as RPTs. However, they will
be shown to be GPTs in~\S\ref{sec_sts2qltl}.
Therefore, the semantics of all atomic RCRS components are GPTs.
This fact, and the closure of GPTs w.r.t. parallel and serial composition
(Theorem~\ref{th:serial_product_gpt}), ensure that we stay within the 
GPT realm as long as no feedback operations are used.
In addition, as we shall prove in Corollary~\ref{component-gpt}, 
components with feedback are also GPTs, as long as they
are deterministic and do not contain {\em algebraic loops}.
An example of a component whose semantics is {\em not} a GPT is:
$$
C = \synfb(\stateless((x,z),(y_1,y_2),y_1=x\land y_2=x))
$$
Then, we have $\sem{C}=\semfb([\sigma_x,\sigma_z \leadsto \sigma_x, \sigma_x])$.
As stated earlier, $\semfb([\sigma_x,\sigma_z \leadsto \sigma_x, \sigma_x])$
is equal to $\{\sigma \leadsto \sigma' \ | \ \true\}$, which is not a GPT
neither an RPT. The problem with $C$ is that it contains an algebraic loop:
the first output $y_1$ of the internal stateless component where feedback
is applied directly depends on its first input $x$. 
Dealing with such components is beyond the scope of this paper, and
we refer the reader to~\cite{PreoteasaTripakisLICS2016}.

\subsubsection{Example: Two Alternative Derivations of the Semantics of Diagram $\mathtt{Sum}$ of Fig.~\ref{fig:diagram}}

To illustrate our semantics, we provide two alternative derivations of the
semantics of the $\mathtt{Sum}$ system of Fig.~\ref{fig:diagram}.

First, let us consider $\mathtt{Sum}$ as a composite component:

$$\mathtt{Sum} = \synfb(\mathtt{Add} \synsop \mathtt{UnitDelay} \synsop \mathtt{Split})$$
where
$$
\begin{array}{lll}
\mathtt{Add}  & = &\determstateless((u,x),\true, u+x)\\
\mathtt{UnitDelay} & = &\determ(x,s,0,\true,x,s)\\
\mathtt{Split}  & = &\determstateless(x,\true, (x,x))\\
\end{array}
$$
We have
$$
\begin{array}{lll}
\sem{\mathtt{Sum}} & = & \sem{\synfb(\mathtt{Add} \synsop \mathtt{UnitDelay} \synsop \mathtt{Split})} \\
	& = & \semfb(\sem{\mathtt{Add}} \circ \sem{\mathtt{UnitDelay}} \circ \sem{\mathtt{Split}} )
\end{array}
$$
For ${\mathtt{Add}}$, ${\mathtt{UnitDelay}}$ and 
${\mathtt{Split}}$, all inputs are legal, so $\illegal_C=\bot$ for
all $C\in\{\mathtt{Add}, \mathtt{UnitDelay},\mathtt{Split}\}$.
After simplifications, we get:
$$
\begin{array}{lll}
\sem{\mathtt{Add}} & = & [\sigma_u,\sigma_x \leadsto \sigma_u + \sigma_x]\\
\sem{\mathtt{UnitDelay}} & = & [\sigma_x \leadsto 0\cdot \sigma_x] \\
\sem{\mathtt{Split}} & = & [\sigma_x \leadsto \sigma_x,\sigma_x]
\end{array}
$$
The semantics of $\mathtt{Sum}$ is given by
$$
\begin{array}{lll}
& \sem{\mathtt{Sum}}\\
=\\
& \semfb(\sem{\mathtt{Add}} \circ \sem{\mathtt{UnitDelay}} \circ \sem{\mathtt{Split}}\\
=\\
& \semfb([\sigma_u,\sigma_x \leadsto \sigma_u + \sigma_x]
  \circ [\sigma_x \leadsto 0\cdot \sigma_x]
  \circ [\sigma_x \leadsto \sigma_x,\sigma_x])\\
  = \\
&\semfb([\sigma_u,\sigma_x \leadsto 
 0\cdot\sigma_x + 0 \cdot \sigma_u,0\cdot\sigma_x + 0 \cdot \sigma_u])\\[1ex]
= &\{\mbox{Using (\ref{eq:ex-feedback}})\} \\[1ex]
& [\sigma_x \leadsto \sigma_y \ |\ \forall i: \sigma_y(i) = \Sigma_{j<i}\sigma_x(j)]

\end{array}
$$
We obtain:
\begin{equation}\label{eq:sem-sum}
\sem{\mathtt{Sum}} = [\sigma_x \leadsto \sigma_y \ |\ \forall i: \sigma_y(i) = \Sigma_{j<i}\sigma_x(j)].
\end{equation}

Next,
let us assume that the system has been characterized already as an
atomic component:

$$\mathtt{SumAtomic} = \sts(x, y, s, s=0, y=s \land s'=s+x).$$

The semantics of $\mathtt{SumAtomic}$ is given by
$$
\sem{\mathtt{SumAtomic}} = \{\neg\illegal_{\mathtt{SumAtomic}}\} 
	\semsop [\sigma_x \leadsto \sigma_y \ |\ 
	\exists \sigma_s : \sigma_s(0) = 0 \land \run_{\mathtt{SumAtomic}}(\sigma_s, \sigma_x,\sigma_y)]
$$
where $\illegal_{\mathtt{SumAtomic}} = \bot$ because there are no restrictions on the inputs
of $\mathtt{SumAtomic}$, and 
$$
\run_{\mathtt{SumAtomic}}(\sigma_s, \sigma_x,\sigma_y) = 
\big(\forall i: \sigma_y(i) = \sigma_s(i) \land \sigma_s(i+1) = \sigma_s(i) + \sigma_x(i) \big)
$$
We have
$$
\sem{\mathtt{SumAtomic}} = [\sigma_x \leadsto \sigma_y \ |\ 
	\exists \sigma_s : \sigma_s(0) = 0 \land \big(\forall i: \sigma_y(i) = \sigma_s(i) \land \sigma_s(i+1) = \sigma_s(i) + \sigma_x(i) \big)]
$$
which is equivalent to (\ref{eq:sem-sum}).

\subsubsection{Characterization of Legal Input Traces}

The following lemma characterizes legal input traces for various types of MPTs:

\begin{lemma} 
The set of legal input traces of an RPT $\{p\} \semsop [r]$ is $p$:
$$\legal(\{p\} \semsop [r]) = p.$$
The set of legal input traces of a GPT $\{r]$ is $\In{r}$:
$$\legal(\{r]) = \In{r}.$$
The set of legal input traces of an STS component $C = \sts(x,y,s,init,r)$ is equal to $\neg \illegal_C$:
$$
\legal(\sem{C}) = \neg \illegal_C.
$$
The set of legal input traces of a QLTL component $C = \qltl(x,y,\varphi)$ is:
$$
\legal(\sem{C}) = \{ \sigma_x \mid \sigma_x \models \exists y: \varphi \}.
$$
\end{lemma}

\subsubsection{Semantic Equivalence and Refinement for Components}
\label{sec_sem_equiv}

\begin{definition} 
\label{def_sem_equiv_refin}
 Two components $C$ and $C'$ are {\em (semantically) equivalent}, denoted $C\equiv C'$, if $\sem{C} = \sem{C'}$.
 Component $C$ is refined by component $C'$, denoted $C\synref C'$,
 if $\sem{C} \semref \sem{C'}$.
\end{definition}
The relation $\equiv$ is an equivalence relation, and $\synref$ is a preorder relation
(i.e., reflexive and transitive). 
We also have 
$$
(C \synref C' \land C' \synref C)
\iff
(C \equiv C')
$$

\subsubsection{Compositionality Properties}
\label{subsec:compositionality}

Several desirable compositionality properties follow from our semantics:

\begin{theorem}\label{th:synrefin}
Let $C_1$, $C_2$, $C_3$, and $C_4$ be four (possibly composite) components. Then:
\begin{enumerate}
\item (Serial composition is associative:) $(C_1 \synsop C_2) \synsop C_3 \equiv C_1 \synsop (C_2 \synsop C_3)$.
\item (Parallel composition is associative:)
	$(C_1 \synpop C_2) \synpop C_3 \equiv C_1 \synpop (C_2 \synpop C_3)$.
\item (Parallel composition distributes over serial composition:) If $\sem{C_1}$ and $\sem{C_2}$ are GPTs
and $\sem{C_3}$ and $\sem{C_4}$ are RPTs, then
$(C_1 \synpop C_2) \synsop (C_3 \synpop C_4) \equiv (C_1 \synsop C_3) \synpop (C_2 \synsop C_4)$.
\item (Refinement is preserved by composition:) If $C_1 \synref C_2$ and $C_3 \synref C_4$, then:
	\begin{enumerate}
		\item $C_1 \synsop C_3 \synref C_2 \synsop C_4$
		\item $C_1 \synpop C_3 \synref C_2 \synpop C_4$
		\item $\synfb(C_1) \synref \synfb(C_2)$
	\end{enumerate}
\end{enumerate}
\end{theorem}

In addition to the above, requirements that a component satisfies are preserved
by refinement. Informally, if $C$ satisfies some requirement $\varphi$ and $C\synref C'$ then $C'$ also satisfies $\varphi$. Although we have not formally defined what requirements are and what it means for a component to satisfy a requirement, these
concepts are naturally captured in the RCRS framework via the semantics of components as MPTs. 
In particular, since our components are generally {\em open} systems
(i.e., they have inputs),
we can express requirements using {\em Hoare triples} of the form
$p\{C\}q$, where $C$ is a component, $p$ is an input property, and $q$ is an output
property. Then, $p\{C\}q$ holds iff the outputs of $C$ are guaranteed to satisfy
$q$ provided the inputs of $C$ satisfy $p$.
Formally: $p \{C\}q \iff p \subseteq \sem{C}(q)$.

\begin{theorem}
\label{thm_substitutability}
 $C\synref C'$ iff $\forall p,q: p\{C\}q \impl p\{C'\}q$.
\end{theorem}
Theorem~\ref{thm_substitutability} shows that refinement is equivalent to {\em substitutability}. Substitutability 
states that a component $C'$ can replace another component $C$ in any context,
i.e., $\forall p,q: p\{C\}q \impl p\{C'\}q$.

%% file: algorithms2.tex

\section{Symbolic Reasoning}
\label{sec:algorithms}

So far we have defined the syntax and semantics of RCRS.
These already allow us to specify and reason about systems 
in a compositional manner.
However, such reasoning is difficult to do ``by hand''. 
For example, if we want to check whether a component $C$ is refined by
another component $C'$, we must resort to proving the refinement relation
$\sem{C} \semref \sem{C'}$ of their corresponding MPTs, $\sem{C}$ and 
$\sem{C'}$. 
This is not an easy task, as MPTs are complex mathematical objects. 
Instead, we would like to have computer-aided, and ideally fully automatic
techniques. In the above example of checking refinement, for instance,
we would like ideally to have an algorithm that takes as input the syntactic
descriptions of $C$ and $C'$ and replies yes/no based on whether 
$\sem{C} \semref \sem{C'}$ holds. We say ``ideally'' because we know that
in general such an algorithm cannot exist. This is because we are not
making a-priori any restrictions on the logics used to describe $C$ and $C'$,
which means that the existence of an algorithm will depend on decidability of
these logics. 
In this section, we describe how reasoning in RCRS can be done
{\em symbolically}, by automatically manipulating the formulas used to specify
the components involved. As we shall show, most problems can be reduced
to checking satisfiability of first-order formulas formed by combinations
of the formulas of the original components. This means that the problems
are decidable whenever the corresponding first-order logics are decidable.

\subsection{Symbolic Transformation of STS Components to QLTL Components}
\label{sec_sts2qltl}

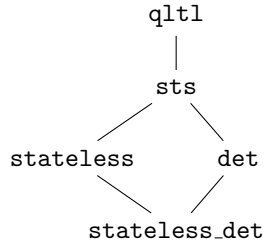
\begin{figure}[h]
\centering
\begin{tikzpicture}
 \node(qltl){$\qltl$};
 \node[below = 3ex of qltl](sts){$\sts$};
 \node[below = 4ex of sts](A){};
 \node[left = 2ex of A](X){$\stateless$};
 \node[right = 2ex of A](Y){$\determ$};
 \node[below = 4ex of A](Z){$\determstateless$};
 \draw[-](sts) -- (qltl);
 \draw[-](sts) -- (X);
 \draw[-](sts) -- (Y);
 \draw[-](X) -- (Z);
 \draw[-](Y) -- (Z);
\end{tikzpicture}
\caption{Lattice of atomic components: lower types are special cases of higher types.}\label{fig:basic}
\end{figure}

Our framework allows the specification of several types of atomic components,
some of which are special cases of others, as summarized in
Fig.~\ref{fig:basic}.
In \S\ref{sec:syntax}, we have already shown how the different types
of STS components are related, from the most specialized deterministic
stateless STS components, to the general STS components.
By definition, the semantics of the special types of STS components is
defined via the semantics of general STS components (see \S\ref{sec:semantics}).
In this subsection, we show that STS components can be viewed as special cases
of QLTL components.

Specifically,
we show how an STS component can be transformed into a semantically equivalent
QLTL component. This transformation also shows that STS property
transformers are a special case of QLTL property transformers, as already
claimed in Fig.~\ref{fig:classification}. Note that
this containment is not obvious simply by looking at the definitions of these
MPT subclasses (c.f.~\S\ref{subsec:comp_semantics}), as QLTL property transformers
are defined as GPTs (equation~\ref{eq_sem_qltl}), whereas STS property
transformers are defined as RPTs (equation~\ref{eq_sem_sts}). 
Although RPTs are generally a superclass, not a subclass, of GPTs, the
transformation proposed below shows that the RPTs obtained from STS components
can indeed be captured as GPTs.
The transformation of
STS into QLTL components also enables us to apply several algorithms which are
available for QLTL components to STS components as well.

Consider an STS component $C = \sts(x, y, s, init, trs)$.
We can transform $C$ into a QLTL component using the mapping $\stsqltl$:
\begin{equation}
\stsqltl(C) = \qltl(x,y, (\forall s, y: init \impl (\varphi \Leads \varphi')) \land 
	(\exists s: init \land \Always \varphi))
\end{equation}
where $\varphi =  trs[s' := \Nextvar s]$
and $\varphi' = (\exists s',y: trs)$.

The theorem that follows demonstrates the correctness of the above transformation:
\begin{theorem}
\label{thm_sts2qltl}
	For any STS component $C=\sts(x, y, s, init, trs)$ s.t.\ $init$ is satisfiable, $C\equiv\stsqltl(C)$.
\end{theorem}

\paragraph{Example.}
It is instructive to see how the above transformation specializes to some
special types of STS components. In particular, we will show how it specializes
to stateless STS components.

Let $C = \stateless(x,y,trs)$ and let $C' = \stateless 2 \sts(C) = 
\sts(x,y,(),\true,trs)$.
Applying the $\stsqltl$ transformation to $C'$, for which $s = ()$ and $init=\true$, we obtain:
$$
\stsqltl(C') = \qltl\big(x,y, (\forall y: (\varphi \Leads \varphi')) \land 
	\Always \varphi)\big)
$$
where $\varphi = trs[():=\Nextvar ()] = trs$ and $\varphi' = (\exists y: trs)$.
Using the properties of Lemma \ref{lem:leads},
and the fact that semantically equivalent LTL formulas result in semantically equivalent QLTL components, we can simplify $\stsqltl(C')$ further:
$$
\begin{array}{ll}
&\stsqltl(C')\\[1ex]
=& \{\mbox{Definition of $\stsqltl$}\}  \\[1ex]
&\qltl\big(x,y, (\forall y: (trs \ \Leads\  (\exists y: trs))) \land 
	\Always\,  trs\big)\\[1ex]
\equiv & \{\mbox{Lemma \ref{lem:leads}, $trs$ does not contain temporal operators, and $y$ is not free
in $(\exists y:trs)$}\}\\[1ex]
&\qltl\big(x,y, ((\exists y:trs) \ \Leads\  (\exists y: trs)) \land 
	\Always\,  trs\big)\\[1ex]
\equiv & \{\mbox{Lemma \ref{lem:leads}}\}\\[1ex]
&\qltl\big(x,y, \Always\, (\exists y:trs) \land 
	\Always\, trs\big)\\[1ex]
\equiv & \{\mbox{Lemma \ref{lem:leads} and $trs$ does not contain temporal operators}\}\\[1ex]
&\qltl\big(x,y, (\exists y: \Always\, trs) \land 
	\Always\,  trs\big)\\[1ex]
\equiv & \{\mbox{Lemma \ref{lem:sem:qltl}}\}\\[1ex]
&\qltl\big(x,y,\, \Always \, trs\big)\\[1ex]
\end{array}
$$
Note that in the above derivation we use the equivalence symbol $\equiv$,
in addition to the equality symbol $=$.
Recall that $\equiv$ stands for semantical equivalence of two components 
(c.f.~\S\ref{sec_sem_equiv}). On the other hand, $=$ for
components means syntactic equality. By definition of the semantics,
if two QLTL formulas are equivalent, then the corresponding QLTL
components are equivalent.

Based on the above, we define the transformation $\stateless 2 \qltl$
of a stateless component $C = \stateless(x,y,trs)$, 
into a QLTL component as follows:
$$
\stateless 2 \qltl(C) = \qltl\big(x,y,\, \Always \, trs\big)
$$

\subsection{Symbolic Transformations of Special Atomic Components to More General Atomic Components}

\begin{sloppypar}
Based on the lattice in Fig.~\ref{fig:basic}, we define all remaining mappings 
from more special atomic components to more general atomic components, by composing
the previously defined mappings $\sts 2\qltl$, $\stateless 2\qltl$, $\stateless 2\sts$, 
$\determ 2\sts$, 
$\determstateless 2\stateless$ and $\determstateless 2\determ$, as appropriately. 
\end{sloppypar}

For mapping stateless deterministic STS components to QLTL components, we have two 
possibilities: 
$\determstateless\to\determ\to\sts\to\qltl$ and $\determstateless \to \stateless \to \qltl$.
We choose the transformation $\determstateless \to \stateless \to \qltl$ because it results
in a simpler formula:
$$
\determstateless 2 \qltl(C) = \stateless 2 \qltl (\determstateless 2 \stateless (C)).
$$

\paragraph{Examples.}

Consider the following STS components:
\begin{eqnarray*}
C_1 & = & \stateless\big(x,y, y>x\big) \\
C_2 & = & \stateless\big(x,(),x>0\big) \\
\mathtt{UnitDelay} &=& \determ\big(x, s, 0, \true, x, s \big)
\end{eqnarray*}
Then:
\begin{eqnarray*}
\stateless 2 \qltl(C_1) & = & \qltl\big(x,y,\,\Always\,y > x\big) \\
\stateless 2 \qltl(C_2) & = & \qltl\big(x,(),\,\Always\,x > 0\big) \\
\determ 2 \qltl(\mathtt{UnitDelay}) &=& \qltl\big(x,y, 
  (\forall s,y: s = 0 \impl \\ 
   &&\quad(\true \land \Nextvar s = x\land y = s)\, \Leads\,
    (\exists s',y: \true\land s'=x \land y = s) )\ \land \\
   &&\quad (\exists s: s = 0 \land \Always\,  (\true \land \Nextvar s = x \land y = s)) \big)
\end{eqnarray*}
Because $(\exists s',y: \true\land s'=x \land y = s) ) \iff \true$, using Lemma \ref{lem:leads} and logical simplifications we obtain:
\begin{eqnarray*}
\determ 2 \qltl(\mathtt{UnitDelay}) &\equiv & \qltl\big(x,y, 
 (\exists s: s = 0 \land \Always\,  (\Nextvar s = x \land y = s)) \big)\\
 &\equiv & \qltl\big(x,y, \,
 (\exists s: s = 0 \land \Always\,  (\Nextvar y = x \land y = s)) \big)\\
 &\equiv & \qltl\big(x,y, \,
 (\exists s: s = 0 \land \Always\,  (y = s) \land \Always\,  (\Nextvar y = x)) \big) \\
 &\equiv & \qltl\big(x,y, \,
 (\exists s: y = 0 \land \Always\,  (y = s) \land \Always\,  (\Nextvar y = x)) \big) \\
 &\equiv & \qltl\big(x,y, \,
 (\exists s: \Always\,  (y = s)) \land  y = 0 \land \Always\,  (\Nextvar y = x) \big) \\
 &\equiv & \qltl\big(x,y, \,
 \Always\,  (\exists s: y = s) \land  y = 0 \land \Always\,  (\Nextvar y = x) \big) \\
 &\equiv & \qltl\big(x,y,\, y = 0 \land \Always\,  (\Nextvar y = x) \big)
\end{eqnarray*}

\subsection{Symbolic Computation of Serial Composition}
\label{sec_symbolic_serial}

Given a composite component $C$ formed as the serial composition of two components $C_1$ and $C_2$, i.e., 
$C = C_1 \synsop C_2$, we would like to compute a new, {\em atomic} component $C_a$, such that $C_a$ is 
semantically equivalent to $C$. Because atomic components are by definition represented syntactically (and symbolically),
being able to reduce composite components into atomic components means that we are able to 
symbolically compute composition operators. 

We start by defining the symbolic serial composition of components
of the same type.

\subsubsection{Symbolic Serial Composition of Two QLTL Components}

 Let $C = \qltl(x,y,\varphi)$ and $C' = \qltl(y,z,\varphi')$ such that
$C \synsop C'$ is well formed.
Then their {\em symbolic serial composition}, denoted $\serial(C,C')$,
is the QLTL component defined by
 
\begin{equation}
\label{eq_serial_compo}
 \serial(C,C') = \qltl\big(x,z, (\forall y: \varphi \impl (\exists z: \varphi')) \land 
      (\exists y:\varphi\land \varphi')\big)
\end{equation}

Note that in the above definition (as well as the ones that follow) we assume that the output variable of $C$ and the input variable of $C'$ have the same name ($y$) and that the names $x$, $y$ and $z$ are distinct. In general, this may not be the case. This is not a problem, as we can always rename variables such that this condition is met.  Note that variable renaming does not change the semantics of components (c.f.~\S\ref{subsubsec:name_scope}). 

The intuition behind the formula in~(\ref{eq_serial_compo}) is as follows.
The second conjunct $\exists y:\varphi\land \varphi'$ ensures that the 
both contracts $\varphi$ and $\varphi'$ of the two components are enforced
in the composite contract. The reason we use $\exists y:\varphi\land \varphi'$
instead of just the conjunction $\varphi\land \varphi'$ is that we want to
eliminate (``hide'') internal variable $y$. 
(Alternatively, we could also have chosen to output $y$ as an additional output, but would then need an additional hiding operator to remove $y$.)
The first conjunct $\forall y: \varphi \impl (\exists z: \varphi')$
is a formula on the input variable $x$ of the composite component
(since all other variables $y$ and $z$ are quantified). This formula
restricts the legal inputs of $C$ to those inputs for which, no matter
which output $C$ produces, this output is guaranteed to be a legal input for
the downstream component $C'$.
For an extensive discussion of the intuition and justification behind this
definition, see~\cite{TripakisLHL11}.

\subsubsection{Symbolic Serial Composition of Two General STS Components}

Let $C = \sts(x,y,s,init,trs)$ and $C' = \sts(y,z,t,init',trs')$
be two general STS components such that $C \synsop C'$ is well formed.
Then:
\begin{equation}
 \serial(C,C') = \sts\big(x,z,(s,t), init\land init', (\exists s',y:trs) \land 
  (\forall s',y: trs \impl (\exists t',z:trs')) 
  \land (\exists y:trs\land trs')\big)
\end{equation}

\subsubsection{Symbolic Serial Composition of Two Stateless STS Components}

Let $C = \stateless(x,y,trs)$ and $C' = \stateless(y,z,trs')$ be two stateless
STS components such that $C \synsop C'$ is well formed. Then
\begin{equation}
 \serial(C,C') = \stateless\big(x,z, (\forall y: trs \impl (\exists z:trs')) 
  \land (\exists y:trs\land trs')\big)
\end{equation}

\subsubsection{Symbolic Serial Composition of Two Deterministic STS Components}

\sloppy

Let $C = \determ(x,s,a,p,next,out)$ and $C' = \determ(y,t,b,p',next',out')$
be two deterministic STS components such that their serial composition $C \synsop C'$ is well formed.
Then:
\begin{equation}
 \serial(C,C') = \determ(x,(s,t), (a,b), p \land p'[y:=out], (next, next'[y:=out]), out'[y:=out])
\end{equation}
where $e[z:=e']$ denotes the substitution of all free occurences of variable $z$
by expression $e'$ in expression $e$.

\subsubsection{Symbolic Serial Composition of Two Stateless Deterministic STS Components}

Finally, let $C = \determstateless(x,p,out)$ and $C' = \determstateless(y,p',out')$
be two stateless deterministic STS components such that their serial composition $C \synsop C'$ is well formed.
Then:
\begin{equation}
 \serial(C,C') = \determstateless(x, p \land p'[y:=out], out'[y:=out])
\end{equation}

\subsubsection{Symbolic Serial Composition of Two Arbitrary Atomic Components}

In general, we define the symbolic serial composition of two atomic components $C$ and $C'$ by using the mappings of less 
general components to more general components (Fig.~\ref{fig:basic}), 
as appropriate.
For example, if $C$ is a deterministic STS component and
$C'$ is a stateless STS component, then $\serial(C,C') = \serial(\determsts(C),
\stateless 2 \sts (C'))$. Similarly,
if $C$ is a QLTL component and $C'$ is a deterministic STS component, then 
$\serial(C,C') = \serial(C,\determ 2 \qltl(C'))$. 
Formally, assume that $\mathit{atm},\mathit{atm}' \in \{\qltl,\sts,\stateless,\determ, \determstateless\}$ are the types of the components 
$C$ and $C'$, and $\mathit{common} = \mathit{atm} \lor \mathit{atm}'$ is 
the least general component type that is more general than 
$\mathit{atm}$ and $\mathit{atm}'$ as defined in 
Fig.~\ref{fig:basic}. Then
$$
  \serial(C,C') = \serial(\mathit{atm}2\mathit{common}(C), 
  \mathit{atm}'2\mathit{common}(C')).
$$

\subsubsection{Correctness of Symbolic Serial Composition}
\label{sec_correctness_symb_serial}

The following theorem demonstrates that our symbolic computations of serial
composition are correct, i.e., that the resulting atomic component 
$\serial(C,C')$ is semantically equivalent to the original composite component
$C \synsop C'$:

\begin{theorem}
\label{thm_symb_serial_correct}
  If $C$ and $C'$ are two atomic components, then
  $$
    C \synsop C' \equiv \serial(C, C').
  $$
\end{theorem}

\paragraph{Examples.}

Consider the following STS components:
\begin{eqnarray*}
C_1 & = & \stateless(u,(x,y),\true) \\
C_2 & = & \determstateless((x,y),y \ne 0, \dfrac{x}{y})
\end{eqnarray*}
Then:
\begin{eqnarray*}
\serial(C_1,C_2) & = & \serial(C_1, \determstateless 2 \stateless(C_2))\\
 & = & \serial(\stateless(u,(x,y),\true), 
 	\stateless((x,y),z,y\ne 0\land z = \dfrac{x}{y})) \\
 & = & \stateless(u, z, (\forall x, y: \true \impl (\exists z: y\ne 0\land z = \dfrac{x}{y})) \\
 && \qquad \land \ (\exists x,y: \true \land y\ne 0\land z = \dfrac{x}{y}))
  ) \\
 & \equiv & \stateless(u,z,\false)
\end{eqnarray*}
As we can see, the composition results in a stateless STS
component with input-output formula $\false$. The semantics
of such a component is $\Fail$, indicating that $C_1$ and $C_2$
are {\em incompatible}.
Indeed, in the case of $C_1\synsop C_2$, the issue is that $C_2$ requires
its second input, $y$, to be non-zero, but $C_1$ cannot guarantee that.
The reason is that the input-output formula of $C_1$ is $\true$, meaning
that, no matter what its input $u$ is, $C_1$ may output {\em any} value
for $x$ and $y$, non-deterministically. This violates the input requirements
of $C_2$, causing an incompatibility.
We will return to this point in~\S\ref{sec_compatibility}.
We also note that this type of incompatibility is impossible to prevent,
by controlling the input $u$. In the example that follows, we see a
case where the two components are {\em not} incompatible, because the
input requirements of the downstream component can be met by strengthening
the input assumptions of the upstream component:

Consider the following QLTL components:
\begin{eqnarray*}
C_3 & = & \qltl(x,y,\Always(x \impl \Evetl y)) \\
C_4 & = & \qltl(y, (), \Always \Evetl y)
\end{eqnarray*}
Then:
\begin{eqnarray*}
\serial(C_3,C_4) & = & \serial(\qltl(x,y,\Always(x \impl \Evetl y)),
  \qltl(y, (), \Always \Evetl y) )\\
  & = & \qltl(x, (), (\forall y : (\Always(x \impl \Evetl y))
  \impl \Always \Evetl y) \land  (\exists y : 
  \Always(x \impl \Evetl y) \land \Always \Evetl y))\\
  & \equiv & \qltl(x, (), \Always \Evetl x)
\end{eqnarray*}
In this example, the downstream component $C_4$ requires its input $y$ to
be infinitely often true ($\Always \Evetl y$).
This can be achieved only if the input $x$ of the upstream component is
infinitely often true, which is the condition derived by the serial
composition of $C_3$ and $C_4$ ($\Always \Evetl x$). Notice that $C_3$
does not impose any a-priori requirements on its input. However, its
input-output relation is the so-called {\em request-response property}
which can be expressed as: {\em whenever the input $x$ is true, the
output $y$ will eventually become true afterwards}
($\Always(x \impl \Evetl y)$). This request-response property implies
that in order for $y$ to be infinitely-often true, $x$ must be infinitely-often
true. Moreover, this is the weakest possible condition that can be enforced
on $x$ in order to guarantee that the condition on $y$ holds.

\subsection{Symbolic Computation of Parallel Composition}
\label{subsec:symbolic_parallel}

Given a composite component formed as the parallel composition of
two components, $C = C_1 \synpop C_2$, we would like to compute an 
atomic component $C_a$, such that $C_a$ is semantically equivalent to $C$.
We show how this can be done in this subsection.
The symbolic computation of parallel composition follows the same pattern as the one for serial composition (\S\ref{sec_symbolic_serial}).

\subsubsection{Symbolic Parallel Composition of Two QLTL Components}

Let $C = \qltl(x,y,\varphi)$ and $C' = \qltl(u,v,\varphi')$.
Then their {\em symbolic parallel composition}, denoted $\Parallel(C,C')$,
is the QLTL component defined by
\begin{equation}
 \Parallel(C, C') = \qltl\big((x,u),(y,v), \varphi \land \varphi'\big)
\end{equation}

As in the case of symbolic serial composition, we assume that 
variable names $x,y,u,v$ are all distinct. If this is not the
case, then we rename variables as appropriately.

\subsubsection{Symbolic Parallel Composition of Two General STS Components}

Let $C = \sts(x,y,s,init,trs)$ and $C' = \sts(u,v,t,init',trs')$.
Then:
\begin{equation}
  \Parallel(C, C') = \sts\big((x,u),(y,v),(s,t), init\land init', trs \land trs'\big)
\end{equation}

\subsubsection{Symbolic Parallel Composition of Two Stateless STS Components}

Let $C = \stateless(x,y,trs)$ and $C' = \stateless(u,v,trs')$. Then
\begin{equation}
 \Parallel(C, C') = \stateless\big((x,u),(y,v), trs \land trs'\big)
\end{equation}

\subsubsection{Symbolic Parallel Composition of Two Deterministic STS Components}

Let $C = \determ(x,s,a,p,next,out)$ and $C' = \determ(u,t,b,p',next',out')$. Then:
\begin{equation}
 \Parallel(C, C') = \determ\big((x,u),(s,t), (a,b), p \land p', (next,next'), (out,out')\big)
\end{equation}

\subsubsection{Symbolic Parallel Composition of Two Stateless Deterministic STS Components}

Let $C = \determstateless(x,p,out)$ and $C' = \determstateless(u,p',out')$. Then:
\begin{equation}
 \Parallel(C, C') = \determstateless\big((x,u), p \land p', (out,out')\big)
\end{equation}

\subsubsection{Symbolic Parallel Composition of Two Arbitrary Atomic Components}

Similar to the symbolic serial composition, we define the symbolic parallel composition 
of two atomic components $C$ and $C'$ by using the mappings of less 
general components to more general components (Fig.~\ref{fig:basic}), 
as appropriate.
Formally, assume that $\mathit{atm},\mathit{atm}' \in \{\qltl,\sts,\stateless,\determ, 
\determstateless\}$ are the types of the components $C$ and $C'$, and 
$\mathit{common} = \mathit{atm} \lor \mathit{atm}'$ is 
the least general component type that is more general than 
$\mathit{atm}$ and $\mathit{atm}'$ as defined in 
Fig.~\ref{fig:basic}. Then
$$
  \Parallel(C,C') = \Parallel(\mathit{atm}2\mathit{common}(C), 
  \mathit{atm}'2\mathit{common}(C')).
$$

\subsubsection{Correctness of Symbolic Parallel Composition}

The following theorem demonstrates that our symbolic computations of parallel
composition are also correct, i.e., that the resulting atomic component 
$\Parallel(C,C')$ is semantically equivalent to the original composite component
$C \synpop C'$:

\begin{theorem}
\label{thm_symb_parallel_correct}
  If $C$ and $C'$ are two atomic components, then
  $$
    C \synpop C' \equiv \Parallel(C, C').
  $$
\end{theorem}

\subsection{Symbolic Computation of Feedback Composition for Decomposable Deterministic STS Components}
\label{sec_algo_feedback}

\subsubsection{Decomposable Components}
\label{sec_decomposability}

We provide a symbolic closed-form expression for the feedback composition
of a deterministic STS component, provided such a component is 
{\em decomposable}.
Intuitively, decomposability captures the fact that the first output of the
component, $y_1$, does not depend on its first input, $x_1$.
This ensures that the feedback composition (which connects $y_1$ to $x_1$)
does not introduce any circular dependencies.

\begin{definition}[Decomposability]
\begin{sloppypar}
Let $C$ be a deterministic STS component
$\determ\big((x_1,\ldots,x_n), s, a, p, next, (e_1, \ldots, e_m)\big)$
or a stateless deterministic STS component 
$\determstateless\big((x_1,\ldots,x_n), p, (e_1, \ldots, e_m)\big)$.
$C$ is
called {\em decomposable} if $x_1$ is not free in $e_1$.
\end{sloppypar}
\end{definition}

\begin{figure}[h]
\centering
\subfloat[]{
\begin{tikzpicture}
 \node[draw, minimum width = 6.5ex, inner ysep = 2ex](A){$e_1$};
 \node[draw, below = 2ex of A, minimum width = 6.5ex](B)
    {$\begin{array}{l}e_2\\ \ldots \\ e_m \end{array}$};
 
 \draw[-latex',thick](B.west) ++ (-9ex, -2ex) coordinate(Y)-- (Y -| B.west);

 \draw[-latex',thick](B.west) ++ (-9ex, -2ex) -- ++(5ex,0) node(X){$\bullet$}
    -- (A.west -| X) -- (A);
    
 \draw[-latex'](B.west) ++ (-9ex, 2ex) coordinate(Z) -- (Z -| B.west);

 \node[draw, fit = (A) (B) (X), inner ysep = 2ex, inner xsep = 2ex](all){};
 
 \node[left = 1ex of Y]{$x_2,\ldots, x_n$};
 \node[left = 1ex of Z]{$x_1$};
 
 \draw[-latex'](A.east) -- ++(5ex, 0);
 \draw[-latex', thick](B.east) -- ++(5ex, 0);
 
\end{tikzpicture}
\label{fig:decomp_orig}
}
$\qquad$
\subfloat[]{
\begin{tikzpicture}
 \node[draw, minimum width = 6.5ex, inner ysep = 2ex](A){$e_1$};
 \node[draw, right = 6ex of A.north east, minimum width = 6.5ex,
 	anchor = north west](B)
    {$\begin{array}{l}e_2\\ \ldots \\ e_m \end{array}$};
 
 \draw[-latex',thick](B.west) ++ (-20ex, -2ex) coordinate(Y)-- (Y -| B.west);

 \draw[-latex',thick](B.west) ++ (-20ex, -2ex) -- ++(5ex,0) node(X){$\bullet$}
    -- (A.west -| X) -- (A);
    
 \draw[-latex'](A.east)  -- ++(3ex,0) node[above]{$x_1$} -- (A -| B.west);

 \node[draw, fit = (A) (B) (X), inner ysep = 2ex, inner xsep = 2ex](all){};
 
 \node[left = 1ex of Y]{$x_2,\ldots, x_n$};
 
 \draw[-latex', thick](B.east) -- ++(5ex, 0);

\end{tikzpicture}
\label{fig:decomp_unfold}
}
\caption{(a) Decomposable deterministic component; (b) the same component after applying feedback, connecting its first output to $x_1$.}
\label{fig:decomposable}
\end{figure}
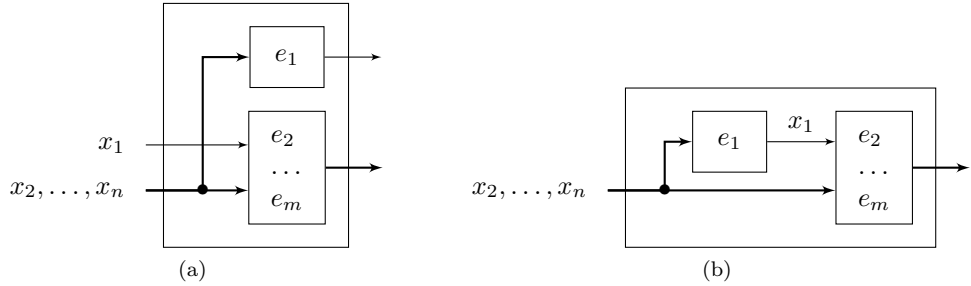

Decomposability is illustrated in Fig.~\ref{fig:decomp_orig}.
The figure shows that expression $e_1$ 
depends only on inputs $x_2,\ldots, x_n$.

\subsubsection{Symbolic Feedback of a Decomposable Deterministic STS Component}

For a decomposable deterministic STS component $C=\determ((x_1,\ldots,x_n), s, a, p, next, (e_1, \ldots, e_m))$, its {\em symbolic feedback composition}, denoted $\symfb(C)$, is the deterministic STS component defined by
\begin{equation}
\symfb(C) = \determ\big((x_2,\ldots,x_n), s, a, p[x_1 := e_1], next[x_1:=e_1], (e_2[x_1:=e_1],\ldots,e_m[x_1:=e_1])\big)
\end{equation}
Thus, computing feedback symbolically consists in removing the first input of
the component and replacing the corresponding variable $x_1$ by the expression
of the first output, $e_1$, everywhere where $x_1$ appears.
The $\symfb$ operator is illustrated in Fig.~\ref{fig:decomp_unfold}.

\subsubsection{Symbolic Feedback of a Decomposable Stateless Deterministic STS Component}

For a decomposable stateless deterministic STS component 
$C=\determstateless((x_1,\ldots,x_n), p, (e_1, \ldots, e_m))$, $\symfb(C)$ is the 
stateless deterministic STS component defined by
\begin{equation}
\symfb(C) = \determstateless\big((x_2,\ldots,x_n), p[x_1 := e_1], (e_2[x_1:=e_1],\ldots,e_m[x_1:=e_1])\big)
\end{equation}

\subsubsection{Correctness of Symbolic Feedback Composition}

\begin{theorem}
\label{thm_symb_feedback_correct}
 If $C$ is a decomposable deterministic STS component, then
 $$
  \synfb(C) \equiv \symfb(C)
 $$
\end{theorem}

Providing closed-form symbolic computations of feedback composition for general components, including possibly non-deterministic STS and QLTL components, is an open problem, beyond the scope of the current paper.
We remark that the straightforward idea of adding to the contract the equality
constraint $x=y$ where $y$ is the output connected in feedback to input $x$,
does not work.\footnote{
One of several problems of this definition is that it does not preserve
refinement. For example, the stateless component with contract $x\ne y$
refines the stateless component with contract $\true$. Adding the constraint
$x=y$ to both contracts yields the components with contracts $x=y$ and $\false$,
respectively, where the latter no longer refines the former. For a more
detailed discussion, see~\cite{PreoteasaTripakisLICS2016}.
}

In fact, even obtaining a semantically consistent compositional definition
of feedback for non-deterministic and non-input-receptive systems is a challenging problem~\cite{PreoteasaTripakisLICS2016}.
Nevertheless, the results that we provide here are sufficient to cover the
majority of cases in practice. 
In particular, the symbolic operator $\symfb$ can be used to handle 
Simulink diagrams, provided these diagrams
do not contain {\em algebraic loops}, i.e., circular dependencies
(see~\S\ref{sec_simplification}).

\subsection{Closure Properties of MPT Subclasses w.r.t. Composition Operators}
\label{sec_closure_MPT_subclasses}

In addition to providing symbolic computation procedures,
the results of the above subsections also prove closure properties
of the various MPT subclasses of RCRS
with respect to the three composition operators.
These closure properties are summarized in Tables~\ref{tab:closure}
and~\ref{tab:fbk:closure}.

In a nutshell, both serial and parallel composition preserve the most
general type of the composed components, according to the lattice
in Fig.~\ref{fig:basic}. For instance, the serial (or parallel)
composition of two stateless STS components is a stateless STS component;
the serial (or parallel) composition of a stateless STS component and a
general STS component is a general STS component; and so on.
Feedback preserves the type of its component (deterministic or stateless
deterministic).

\begin{table}[h]
\centering
\begin{tabular}{c | c | c | c | c | c}
$\synsop$ and $\synpop$ & $\qltl$ & $\sts$ & $\stateless$ & $\determ$ & $\determstateless$ \\
\hline
$\qltl$ & $\qltl$ & $\qltl$ & $\qltl$ & $\qltl$ & $\qltl$ \\
\hline
$\sts$  & $\qltl$ & $\sts$ & $\sts$ & $\sts$ & $\sts$ \\
\hline  
$\stateless$ & $\qltl$ & $\sts$ & $\stateless$ & $\sts$ & $\stateless$ \\
\hline
$\determ$  & $\qltl$ & $\sts$ & $\sts$ & $\determ$ & $\determ$  \\
\hline
$\determstateless$ & $\qltl$ & $\sts$ & $\stateless$ & $\determ$ & $\determstateless$  \\
\hline
\end{tabular}
\caption{Closure properties of serial and parallel \label{tab:closure} compositions. The table is to be read as follows: given atomic components $C_1,C_2$ of types as specified in a row/column pair, the serial or parallel composition of $C_1$ and $C_2$ is an atomic component of type as specified in the corresponding table entry.}
\end{table}

\begin{table}[h]
\centering
\begin{tabular}{c | c | c }
$\synfb$ &  $\determ$ and decomposable & $\determstateless$ and decomposable \\
\hline
&  $\determ$ & $\determstateless$ 
\end{tabular}
\caption{Closure properties of feedback \label{tab:fbk:closure} composition.}
\end{table}

\subsection{Symbolic Simplification of Arbitrary Composite Components}
\label{sec_simplification}

The results of the previous subsections show how to simplify into an atomic
component the serial or parallel composition of two atomic components, or
the feedback composition of an atomic decomposable component. We can combine these
techniques in order to provide a general symbolic simplification algorithm:
the algorithm takes as input an arbitrarily complex composite component,
and returns an equivalent atomic component. The algorithm is shown
in Fig.~\ref{fig_atomic_algo}.

\begin{figure}[h]
\centering
$
\begin{array}{lll}
\symbolic(C):\\[1ex]
\qquad \mathsf{if} \ C \ \mbox{is atomic } \mathsf{then} \\[1ex]
\qquad \qquad \mathsf{return} \ C \\[1ex]
\qquad \mathsf{else \ if} \ C \ \mbox{is } C' \synsop C'' \ \mathsf{then} \\[1ex]
\qquad \qquad \mathsf{return} \ \serial(\symbolic(C'),\symbolic(C'')) \\[1ex]
\qquad \mathsf{else \ if} \ C \ \mbox{is } C' \synpop C'' \ \mathsf{then} \\[1ex]
\qquad \qquad \mathsf{return} \ \Parallel(\symbolic(C'),\symbolic(C'')) \\[1ex]
\qquad \mathsf{else \ if } \ C \mbox{ is } \synfb(C') \ \mathsf{then} \\[1ex]
\qquad \qquad C'' := \symbolic(C')\\[1ex]
\qquad \qquad \mathsf{if } \ C'' \mbox{ is decomposable} \ \mathsf{then} \\[1ex]
\qquad \qquad \qquad \mathsf{return} \ \symfb(C'') \\[1ex]
\qquad \qquad  \mathsf{else} \\
\qquad \qquad \qquad \mathsf{fail} \\
\qquad \mathsf{else} \mbox{ /* impossible by definition of syntax */} \\
\qquad \qquad \mathsf{fail}
\end{array}
$
\caption{Simplification algorithm for arbitrary composite components.}
\label{fig_atomic_algo}
\end{figure}

The algorithm fails only in case it encounters the feedback of a
non-decomposable component.
Recall that decomposability implies determinism (c.f.~\S\ref{sec_decomposability}), which means that the test {\em $C''$ is decomposable} means that $C''$ is of
the form
   $\determ((x_1,\ldots), s, a, p, next, (e_1, \ldots))$
   or $\determstateless((x_1,\ldots), p, (e_1, \ldots))$
   and $x_1$ is not free in $e_1$.
   
We note that in practice, our RCRS implementation on top of Isabelle
performs more simplifications in addition to those 
performed by the procedure $\symbolic$.
For instance, our implementation may be able to simplify a logical formula
$\phi$ into an equivalent but simpler formula $\phi'$ (e.g., by eliminating
quantifiers from $\phi$), and consequently also simplify a component, say, $\qltl(x,y,\phi)$ into an equivalent but simpler component
$\qltl(x,y,\phi')$.
These simplifications very much depend on the logic used in the components.
Describing the simplifications that our implementation performs
is outside the scope of the current paper,
as it belongs in the realm of computational logic.
It suffices to say that our tool is not optimized for this purpose,
and could leverage specialized tools and relevant advances in the field
of computational logic.

\subsubsection{Deterministic and Algebraic Loop Free Composite Components}

In order to state and prove correctness of the algorithm, we extend the notion of
determinism to a composite component. 
We also introduce the notion of {\em algebraic loop free} components, which capture systems
with no circular and intantaneous input-output dependencies.

A (possibly composite) component $C$ is said to be {\em deterministic} if every atomic component of $C$ is either a deterministic STS component or a stateless deterministic STS component.
Formally, $C$ is deterministic iff $\deterministic(C)$ is true, where
$\deterministic(C)$ is defined inductively on the structure of $C$:
$$
\begin{array}{lll}
\deterministic(\determ(x, s, a, p, n, e)) & = & \true\\
\deterministic(\determstateless(x, p, e)) & = & \true\\
\deterministic(\sts(x,y,s, init, trs)) & = & \false\\
\deterministic(\stateless(x,y,trs)) & = & \false\\
\deterministic(C\synsop C') & = & \deterministic(C) \land \deterministic(C') \\
\deterministic(C\synpop C') & = & \deterministic(C) \land \deterministic(C') \\
\deterministic(\synfb(C)) &=& \deterministic(C)
\end{array}
$$
Notice that this notion of determinism applies to a generally {\em composite} component $C$, i.e., a syntactic term in our algebra of components, involving atomic components possibly composed via the three composition operators.  This notion of determinism is the generalization of the syntactically deterministic STS components, which are atomic.  This notion of determinism is also distinct
from any {\em semantic} notion of determinism (which we have not
introduced at all in this paper, as it is not needed). 

For a deterministic component we define its {\em output input dependency relation}. Let $C$ be deterministic, and let
$\intype(C) = X_1\times\ldots\times X_n$ and $\outtype(C) = Y_1 \times \ldots \times Y_m$. 
The relation $\OI(C)\subseteq\{1,\ldots,m\}\times\{1,\ldots,n\}$ 
is defined inductively on the structure of $C$:
$$
\begin{array}{llll}
\OI(\determ((x_1,\ldots,x_n), s, a, p, next, (e_1,\ldots,e_m))) & = & \{(i,j)\ |\ x_j \mbox{ is free in } e_i\} \\
\OI(\determstateless((x_1,\ldots,x_n), p, (e_1,\ldots,e_m))) & = & \{(i,j)\ |\ x_j \mbox{ is free in } e_i\} \\
\OI(C\synsop C') & = & \OI(C') \circ \OI(C) \\
\OI(C\synpop C') & = & \OI(C) \cup \{(i+m,j+n)\ | \ (i,j) \in \OI(C')\} \\
& & \text{ where } \intype(C) = X_1\times\ldots\times X_n \\
& & \text{ and } \outtype(C) = Y_1 \times \ldots \times Y_m \\
\OI(\synfb(C)) &=& \{(i,j)\ |\ i > 0 \land j > 0 \land 
   ((i + 1,j + 1) \in \OI(C) \\ && \ \ \ \ \lor\ ((i + 1, 1)\in \OI(C) \land (1, j+1)\in \OI(C)))\}
\end{array}
$$
The intuition is that $(i,j)\in\OI(C)$ iff the $i$-th output of $C$ depends
on its $j$-th input.

The $\OI$ relation is preserved by the symbolic operations, as shown by the
following lemma:

\begin{lemma}\label{lemma:OI}
If $C$ and $C'$ are deterministic STS components, then 
\begin{eqnarray*}
\OI(C\synsop C') & = & \OI(\serial(C,C')) \\
\OI(C\synpop C') & = & \OI(\Parallel(C,C')).
\end{eqnarray*}
If $C$ is also decomposable, then
\begin{eqnarray*}
\OI(\synfb(C)) & = & \OI(\symfb(C)).
\end{eqnarray*}
\end{lemma}

We introduce now the the notion of algebraic loop free component.
Intuitively, a (possibly composite) deterministic component $C$ is algebraic loop free if, whenever $C$ contains a subterm of the form $\synfb(C')$, the first output of $C'$ does not depend on its first input. This implies that whenever a feedback
connection is formed, no circular dependency is introduced. It also ensures that
the simplification algorithm will never fail.
Formally, for  a component $C$ such that 
$\deterministic(C)$ is true, $\decomp(C)$ is defined inductively on the structure of $C$:

$$
\begin{array}{lll}
  \decomp(C) & = & \true \mbox{ if $C=\determ(x,s,a,p,next,out)$ 
  } \\
  \decomp(C) & = & \true \mbox{ if $C=\determstateless(x,p,out)$ 
  }\\
  \decomp(C\synsop C') & = & \decomp(C) \land \decomp(C')\\
  \decomp(C\synpop C') & = & \decomp(C) \land \decomp(C')\\
  \decomp(\synfb(C)) & = & \decomp(C) \land (1,1) \not \in \OI(C) \\
\end{array}
$$

\subsubsection{Correctness of the Simplification Algorithm}

\begin{theorem}\label{thm:atomic}
Let $C$ be a (possibly composite) component.
\begin{enumerate}
\item
If $C$ does not contain any $\synfb$ operators then $\symbolic(C)$ 
does not fail and returns an atomic component such that $\symbolic(C) \equiv C$.
\item
 If $\deterministic(C)\land\decomp(C)$ is true then $\symbolic(C)$ 
does not fail and returns an atomic component such that $\symbolic(C) \equiv C$.
Moreover, $\symbolic(C)$ is a deterministic STS component and
 $$
 \OI(C) = \OI(\symbolic(C)).
 $$
\end{enumerate}
\end{theorem}
\proof
The first part of the theorem is a consequence of the fact that the
symbolic serial and parallel compositions are defined for all
atomic components and return equivalent atomic components,
by Theorems~\ref{thm_symb_serial_correct} and~\ref{thm_symb_parallel_correct}.

For the second part, since we have a recursive procedure, we prove its correctness by asumming
the correctness of the recursive calls. Additionally, the termination of
this procedure is ensured by the fact that all recursive calls are
made on ``smaller" components. Specifically:
we assume that both $\deterministic(C)$ and $\decomp(C)$ hold; and we prove
that $\symbolic(C)$ does not fail, $\symbolic(C)$ is a deterministic
STS component, $\symbolic(C)\equiv C$, and $\OI(C) = \OI(\symbolic(C))$.

We only consider the case $C = \synfb(C')$. All other cases are similar, but simpler.
Because $\deterministic(C)$ and $\decomp(C)$ hold, we have that
$\deterministic(C')$ and $\decomp(C')$ also hold, and in addition
$(1,1) \not \in \OI(C')$. Using 
the correctness  assumption for the recursive call we have that 
$\symbolic(C')$ does not fail, $C'' = \symbolic(C')$ is a deterministic 
STS component, $C'' \equiv C'$, and $\OI(C'') = \OI(C')$.

Because $C''$ is a deterministic STS component and $(1,1)\not\in \OI(C') = \OI(C'')$, 
$C''$ is decomposable. From this we have that
$D := \symfb(C'')$ is defined. Therefore, $\symbolic(C)$ returns $D$ and
does not fail.
It remains to show that $D$ has the desired properties.
By the definition of $\symfb(C'')$ and the fact that $C''$ is a
decomposable deterministic STS component, $D$ is also a
deterministic STS component.
We also have:
\begin{eqnarray*}
D  =  \symfb(C'') \equiv \synfb(C'') \equiv \synfb(C') = C
\end{eqnarray*}
where $\symfb(C'') \equiv \synfb(C'')$ follows from
Theorem~\ref{thm_symb_feedback_correct} and
 $\synfb(C'') \equiv \synfb(C')$ follows from $C'' \equiv C'$ and the
semantics of $\synfb$.

Finally, using Lemma \ref{lemma:OI} and $ \OI(C'') = \OI(C')$, we have
$$
\OI(D)  =  \OI(\symfb(C''))  = \OI(\synfb(C'')) = \OI(\synfb(C')) = \OI(C)
$$
\qed

\begin{corollary}\label{component-gpt}
If a component $C$ does not contain any $\synfb$ operators or if $\deterministic(C)\land\decomp(C)$ is true, then
$\sem{C}$ is a GPT.
\end{corollary}

Note that condition $\deterministic(C)\land\decomp(C)$ is sufficient, but
not necessary, for $\sem{C}$ to be a GPT.
For example, consider the following components:
$$
  \begin{array}{lll}
  \mathtt{Const}_\false &=& \determstateless\big(x, \true, \false\big) \\[1ex]
  \mathtt{And} &=& \determstateless\big((x,y), \true, (x\land y,x\land y)\big) \\[1ex]
  C &=& \mathtt{Const}_\false\synsop\synfb(\mathtt{And})
  \end{array}
$$
$\mathtt{Const}_\false$ outputs the constant $\false$.
$\mathtt{And}$ is a version of logical and with two identical outputs
(we need two copies of the output, because one will be eliminated once we
apply feedback).
$C$ is a composite component, formed by first connecting the first
output of $\mathtt{And}$ in feedback to its first input, and then
connecting the output of $\mathtt{Const}_\false$ to the second input
of $\mathtt{And}$ (in reality, to the only remaining input of $\synfb(\mathtt{And})$).
Observe that $C$ has algebraic loops, that is, $\decomp(C)$ does not hold.
Yet it can be shown that $\sem{C}$ is a GPT (in particular,
we can show that $C\equiv\mathtt{Const}_\false$).

\paragraph{Example.}

The simplification algorithm applied to the component from Fig.~\ref{fig:diagram}
results in
$$
\symbolic(\mathtt{Sum}) = 
\symbolic(\synfb(\mathtt{Add} \synsop \mathtt{UnitDelay} \synsop \mathtt{Split}))
= \determ(y,s,0,s+y,s).
$$
To see how the above is derived, let us first calculate $\symbolic(\mathtt{Add} \synsop \mathtt{UnitDelay} \synsop \mathtt{Split})$:
$$
\begin{array}{lll}
&\symbolic(\mathtt{Add} \synsop \mathtt{UnitDelay} \synsop \mathtt{Split})\\
=\\
&\serial(\symbolic(\mathtt{Add} \synsop \mathtt{UnitDelay}),\ \mathtt{Split})\\
=\\
&\serial(\serial(\mathtt{Add},\  \mathtt{UnitDelay}),\ \mathtt{Split})\\[1ex]
=&\{\mbox{Expanding $\mathtt{Add}$ and  $\mathtt{UnitDelay}$ and choosing suitable variable names}
\}\\[1ex]
&\serial(\serial(\determstateless((x,y),\true,x + y),\  \determ(z,s,0,\true,z,s)),\ \mathtt{Split})
\\[1ex]
=&\{\mbox{Computing inner $\serial$}\}\\[1ex]
&\serial(\determ((x,y),s,0,\true,x + y,s),\ \mathtt{Split})\\[1ex]
=&\{\mbox{Expanding $\mathtt{Split}$ and choosing suitable variable names}
\}\\[1ex]
&\serial(\determ((x,y),s,0,\true,x + y,s),\ \determstateless(z,\true,(z,z)))\\[1ex]
=&\{\mbox{Computing remaining $\serial$}\}\\[1ex]
&\determ((x,y),s,0,\true,x + y,(s,s))
\end{array}
$$
We calculate now $\symbolic(\synfb(\determ((x,y),s,0,\true,x + y,(s,s))))$:
$$
\begin{array}{lll}
&\symbolic(\synfb(\determ((x,y),s,0,\true,x + y,(s,s))))\\[1ex]
= & \{ \mbox{$x$ is not free in $s$}\} \\[1ex]
& \symfb(\determ((x,y),s,0,\true,x + y,(s,s)))\\[1ex]
= & \{ \mbox{Computing $\symfb$}\} \\[1ex]
& \determ(y,s,0,\true,s + y, s)

\end{array}
$$

\subsection{Checking Validity and Compatibility}
\label{sec_compatibility}

Recall the example given in~\S\ref{sec_correctness_symb_serial}, of the
serial composition of components $C_1$ and $C_2$, resulting in a component
with input-output relation $\false$, implying that $\sem{C_1\synsop C_2}=\Fail$.
When this occurs, we say that $C_1$ and $C_2$ are {\em incompatible}.
We would like to catch such incompatibilities. This amounts to first
simplifying the serial composition $C_1\synsop C_2$ into an atomic component
$C$, and then checking whether $\sem{C}=\Fail$.

In general, we say that a component $C$ is {\em valid} if $\sem{C} \neq \Fail$.
Given a component $C$, we can check whether it is valid, as follows. 
First, we simplify $C$ to obtain an atomic component $C'=\symbolic(C)$.
If $C'$ is a QLTL component of the form $\qltl(x,y,\varphi)$ then $C'$
is valid iff $\varphi$ is satisfiable.
The same is true if $C'$ is a stateless STS component of the form $\stateless(x,y,\varphi)$.
If $C'$ is a general STS component then we can first transform it into a QLTL
component and check satisfiability of the resulting QLTL formula.

\begin{theorem}
If $C$ is an atomic component of the form $\qltl(x,y,\varphi)$ or
$\stateless(x,y,\varphi)$, then $\sem{C} \neq \Fail$ iff $\varphi$ is satisfiable.
\end{theorem}

As mentioned in the introduction,
one of the goals of RCRS is to function as a behavioral type system
for reactive system modeling frameworks such as Simulink. 
In the RCRS setting, type checking consists in checking properties such as
compatibility of components, as in the example $C_1\synsop C_2$
of~\S\ref{sec_correctness_symb_serial}.
When components are compatible,
by computing new (stronger) input preconditions like in the example
$C_3\synsop C_4$ of~\S\ref{sec_correctness_symb_serial}, RCRS can
be seen as enabling to perform behavioral type inference.
Indeed, the new derived condition $\Always \Evetl x$ in the above example
can be seen as an inferred type of the composite component $C_3\synsop C_4$.

We note that the decidability of the satisfiability question for $\varphi$
depends on the logic used and the domains of the variables in the formula.
For instance, although $\varphi$ can be a QLTL formula, if it restricts
the set of constants to $\true$, has no functional symbols, and only
equality as predicate symbol, then it is equivalent to a QPTL formula,\footnote{
For example, the atomic QLTL formula $\Nextvar \Nextvar x = \Nextvar y$
can be translated into the LTL formula $\Next \Next x \eqv \Next y$, 
and the formula $\Nextvar \Nextvar \Nextvar x = \true$ into $\Next \Next \Next x$.
}
for which we can use available techniques~\cite{SistlaVardiWolper87}.

\subsection{Checking Input-Receptiveness and Computing Legal Inputs Symbolically}

Given a component $C$, we often want to check whether it is {\em input-receptive}, i.e., whether $\legal(\sem{C}) = \top$, or equivalently, $\sem{C}(\top)=\top$. More generally, we may want to compute the legal input values for $C$,
which is akin to type inference as discussed above.
To do this, we will provide a symbolic method to compute $\legal(\sem{C})$
as a formula $\symblegal(C)$.
Then, checking that $C$ is input-receptive amounts to checking that the
formula $\symblegal(C)$ is valid, or equivalently, checking that
$\neg\symblegal(C)$ is unsatisfiable.
We assume that $C$ is atomic (otherwise, we first simplify
$C$ using the algorithm of \S\ref{sec_simplification}).

\begin{definition}
Given an atomic component $C$, we define $\symblegal(C)$,
a formula characterizing the legal inputs of $C$.
$\symblegal(C)$ is defined based on the type of $C$:
\begin{eqnarray}
  \symblegal\big( \qltl(x,y,\varphi) \big) &\!\!\!=\!\!\!& (\exists y : \varphi) \\
  \symblegal\big( \sts(x,y,s,init,trs) \big) &\!\!\!=\!\!\!& (\forall s, y: init \impl (\varphi \Leads \varphi')) \\
  \symblegal\big( \stateless(x,y,trs) \big) &\!\!\!=\!\!\!& \Always (\exists y: trs) \\
  \symblegal\big( \determ(x,s,a,p,next,out) \big) &\!\!\!=\!\!\!& 
  		(\forall s,y: s = a \impl (\Nextvar s = next \land y = out) \Leads p) \\
  \symblegal\big( \determstateless(x,p,out) \big) &\!\!\!=\!\!\! & \Always p 
  		\delete{
\mbox{For a deterministic STS component $C$, } 
		\symblegal(C) &\!\!\!=\!\!\!& \symblegal(\determ 2\sts(C)) \\
\mbox{For a stateless deterministic STS component $C$, } 
		\symblegal(C) &\!\!\!=\!\!\!& \symblegal(\determstateless 2\stateless(C))
		}
\end{eqnarray}
where $\varphi =  trs[s' := \Nextvar s]$ 
and $\varphi' = (\exists s',y: trs)$.
\end{definition}

The next theorem shows that $\symblegal$ correctly characterizes the semantic
predicate $\legal$:
\begin{theorem}
\label{thm_legal}
 If $C$ is an atomic component , then
 $$
  \legal(\sem{C}) = \{\sigma_x\ | \ \sigma_x \models \symblegal(C)\}.
 $$
\end{theorem}

It follows from Theorem~\ref{thm_legal} that a component $C$ is input-receptive
iff the formula $\symblegal(C)$ is valid.

\delete{
Next theorem introduces neccessary and suffucient conditions for
STS components to be input complete:
\begin{theorem}
The STS component $C = \sts(x,y,s,init,trs)$ is input complete
(i.e. $\symblegal(C)$ is valid) if and only if the following formulas are valid:
$$
\begin{array}{l}
init \impl p\\[1ex]
(\exists t : init[s:=t] \land ((\exists x, y: trs[s,s':=t,\Nextvar t]) \Until t = s)) \land p \land trs \impl p[s,x:=s',x']
\end{array}
$$
where $p = (\exists s',y: trs)$, and $t$ and $x'$ are new variable names.
\end{theorem} 
The QLTL formula $(\exists t : init[s:=t] \land ((\exists x, y: trs[s,s':=t,\Nextvar t]) \Until t = s))$ contains only variable
$s$ free and expresses the fact that that $s$ is reachable from
some state satisfying the formula $init$.
}

\subsection{Checking Refinement Symbolically}

We end this section by showing how to check whether a component refines
another component. Again, we will assume that the components in question
are atomic (if not, they can be simplified using the $\symbolic$ procedure).

\begin{theorem}
\label{th:refin}
Let $C_1 = \sts(x,y,s,init,r_1)$, $C_1'=\sts(x,y,s,init',r_1')$,
$C_2 = \stateless(x,y,r_2)$, $C_2' = \stateless(x,y,r_2')$,
$C_3=\qltl(x,y,\varphi)$, and $C_3' = \qltl(x,y,\varphi')$.
Then:
\begin{enumerate}
  \item $C_1$ is refined by $C_1'$ if
    the formula
    $$ (init' \impl init) \land ((\exists s',y : r_1) \impl (\exists s',y : r_1')) \land 
    ((\exists s',y : r_1) \land r_1' \impl r) $$
    is valid.
  \item $C_2$ is refined by $C_2'$ if and only if the formula
    $$
      \big((\exists y : r_2) \impl (\exists y : r_2')\big) \land 
      \big((\exists y : r_2) \land r_2' \impl r_2\big)
    $$
    is valid.
  \item $C_3$ is refined by $C_3'$ if and only if the formula
    $$
      \big((\exists y : \varphi) \impl (\exists y: \varphi')\big) \land \big(((\exists y : \varphi) \land \varphi') \impl \varphi\big)
    $$
    is valid.
\end{enumerate}
\end{theorem}

As the above theorem shows, checking refinement amounts to checking validity
(or equivalently, satisfiability of the negation) of first-order formulas
formed by the various symbolic expressions in the component specifications.
The exact logic of these formulas depends on the logics used by the components.
For example, if $C_3$ and $C_3'$ both use quantifier-free LTL for $\phi$ and
$\phi'$, then in order to check refinement we need to check satisfiability
of a first-order QLTL formula.

\paragraph{Examples.}
Consider again the QLTL component $C = \qltl((),t,\mathtt{oven})$, introduced in \S\ref{subsec:qltl}, where 
\begin{eqnarray*}
\mathtt{oven} &=& (t = 20 \land ((t < \Nextvar t \land t < 180) \Until \mathtt{thermostat}))\\
\mathtt{thermostat}& =& \Always (180 \leq t \land t \leq 220)
\end{eqnarray*}
Let us introduce a refined version $C'$ of $C$:
$$
\begin{array}{lll}
C' & = & \sts((),\ t,\ (s,sw), \ \mathtt{init}, \ \mathtt{trs}) \mbox{\ where }\\[1ex]
\mathtt{init} &= & s = 20 \land sw = \mathsf{on} \\[1ex]
\mathsf{trs}& = & (t = s) \ \land \\[1ex] && (\mathtt{if\ }sw=\mathsf{on} \mathtt{\ then\ } s < s' < s + 5 \mathtt{\ else\ (if \ } 
s > 10 \mathsf {\ then \ }s - 5 < s' < s \mathsf {\ else \ 
} s' = s)) \  \land \\[1ex]
&&  (\mathtt{if\ }sw = \mathsf{on} \land s > 210 \mathtt{\ then\ } sw' = \mathsf{off} \mathtt{\ else}\\[1ex]
&&  \qquad (\mathtt{if \ } sw = \mathsf{off} \land s < 190 \mathtt{\ then\ } sw' = \mathsf{on} 
  \mathtt{\ else \ }sw' = sw)) 
\end{array}
$$
$C'$ is an STS component with no input variables, output variable $t$, and state
variables $s$ and $sw$, recording
the current temperature of the oven, and the on/off status of the switch,
respectively.
When $sw$ is $\mathsf{on}$, the temperature increases nondeterministically by up to $5$ units, otherwise the temperature decreases nondeterministically by up to $5$ units.
When the temperature exceeds 210, the switch is turned $\mathsf{off}$;
when the temperature is below $190$, the switch is turned $\mathsf{on}$;
otherwise $sw$ remains unchanged. The output $t$ is always equal to the current
state $s$.
Initially the temperature is $20$, and $sw$ is $\mathsf{on}$.

Using Theorem~\ref{th:refin}, and the properties of $\sts 2 \qltl$ we have:
$$
\begin{array}{lll}
& C \synref C' \\
\iff \\
& C \synref \sts 2 \qltl(C')\\
\iff \\
& \qltl((),t,\mathtt{oven}) \synref \qltl((),t, (\forall s,sw,t:\mathtt{init} \impl 
  (\varphi \Leads \varphi')) \land (\exists s,sw: \mathtt{init} \land \Always \varphi)) \\[1ex]
  & \mbox{\ \ \ \ where } \varphi = \mathtt{trs}[s',sw':=\Nextvar s,\Nextvar sw] \mbox{ and }
   \varphi' = (\exists s',sw',t:\mathtt{trs}) \\[1ex]
 
\iff & \{\mbox{Using Lemma~\ref{lem:leads}, because } \varphi' \iff \true\}\\[1ex]
 
& \qltl((),t,\mathtt{oven}) \synref \qltl((),t, (\exists s,sw: \mathtt{init} \land \Always \varphi)) \\[1ex]

\iff & \{\mbox{Using Theorem~\ref{th:refin}}\}\\[1ex]
&  \big((\exists t: \mathtt{oven} ) \impl (\exists t, s,sw: \mathtt{init} \land \Always \varphi)\big) \land \big(((\exists t: \mathtt{oven}) \land (\exists s,sw: \mathtt{init} \land \Always \varphi)) \impl \mathtt{oven}\big)
\mbox{ is valid}
\\[1ex]

\iff & \{\mbox{Because } (\exists t: \mathtt{oven} ) \iff \true \mbox{ and } (\exists t, s,sw: \mathtt{init} \land \Always \varphi) \iff \true\}\\[1ex]

& \big((\exists s,sw: \mathtt{init} \land \Always \varphi) \impl \mathtt{oven}\big)
\mbox{ is valid} 


\end{array}
$$
Thus, checking whether $C'$ refines $C$ amounts to checking whether the QLTL
formula $\big((\exists s,sw: \mathtt{init} \land \Always \varphi) \impl \mathtt{oven}\big)$ is valid.
This indeed holds for this example and can be shown using logical reasoning.

The above example is relatively simple in the sense that in the end
refinement reduces to checking implication between the corresponding
contracts. Indeed, this is always the case for input-receptive systems,
as in the example above. However, refinement is {\em not} equivalent
to implication in the general case of non-input-receptive systems.
For example, consider the components:
\begin{eqnarray*}
C_1 &=& \stateless(x,y,x\ge 0 \land y\ge x) \\
C_2 & =& \stateless(x,y, x\le y\le x+10) 
\end{eqnarray*}
Using Theorem~\ref{th:refin}, we have:
$$
\begin{array}{lll}
& C_1 \synref C_2 \\[1ex]
\iff & \{\mbox{Theorem~\ref{th:refin}}\}\\[1ex]
& ((\exists y: x \ge 0 \land y \ge x) \impl (\exists y: x \le y \le x + 10)) \ \land  \\[1ex] & \qquad 
 ((\exists y: x \ge 0 \land y \ge x) \land x \le y \le x + 10
\impl  x \ge 0 \land y \ge x) \mbox{ is valid}\\[1ex]
\iff & \{\mbox{Arithmetic and logical reasoning}\}\\[1ex]
& \true
\end{array}
$$

Note that the second and third parts of Theorem~\ref{th:refin} provide necessary {\em and}
sufficient conditions, while the first part only provides a sufficient, but
generally not necessary condition. Indeed, the condition is generally
not necessary in the case of STS components with state, as state space
computation is ignored by the condition. This can be remedied by
transforming STS components into equivalent QLTL components and then
applying the second part of the theorem. An alternative which may be
more tractable, particularly in the case of finite-state systems,
is to use techniques akin to strategy synthesis in games, such
as those proposed in~\cite{TripakisLHL11}
for finite-state relational interfaces.

Another limitation of the first part of Theorem~\ref{th:refin} is 
that it requires the two STS components to have the same state space,
i.e., the same state variable $s$. This restriction can be lifted
using the well-known idea of data refinement~\cite{hoare:1972,back-1980}.

\begin{theorem}
Let $C_1 = \sts(x,y,s,init,r)$, $C_2=\sts(x,y,t,init',r')$ be two STS components, and
$D$ a (data refinement) expression on variables $s$ and $t$. 
Let $p = (\exists s',y: r)$ and $p' = (\exists t',y: r')$.
If the formulas
$$
\begin{array}{ll}
(\forall t: init' \impl (\exists s : D \land init))\\[1ex]
(\forall t, x, s : D \land p \impl p') \\[1ex]
(\forall t, x, s, t', y : D \land p \land  r' \impl (\exists s' : D[t,s:=t',s'] \land r))
\end{array}
$$
are valid, then $C_1$ is refined by $C_2$. 
\end{theorem}

%% file: implementation.tex
\section{Toolset and Case Studies}
\label{sec_implementation}



\begin{figure}
 \centering
\ifdefined\JACM
 \includegraphics[scale=0.60]{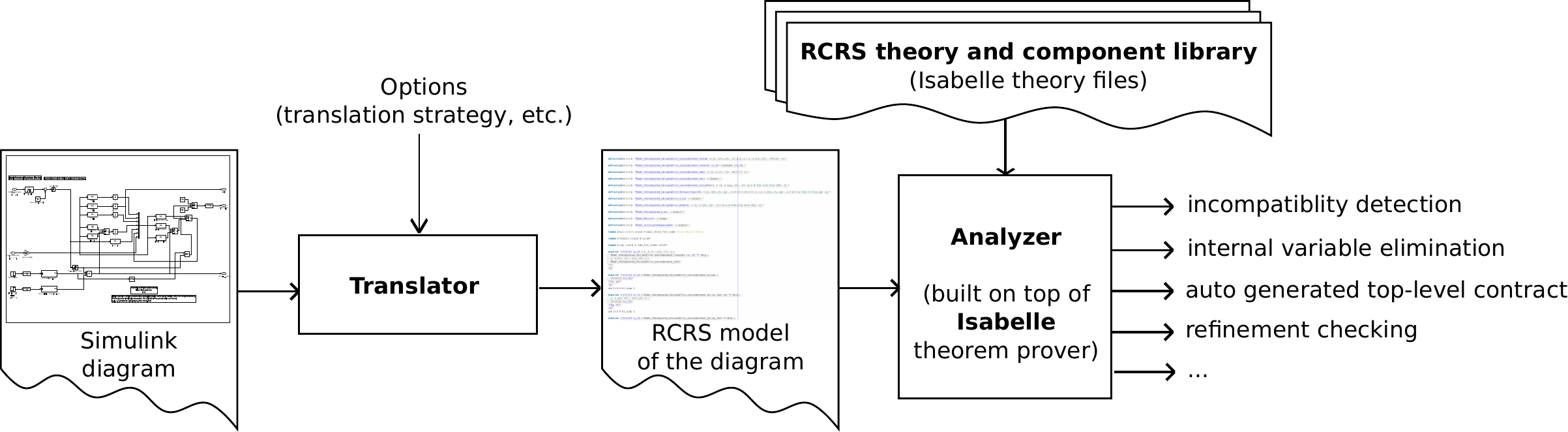}
\else
 \includegraphics[scale=0.65]{toolset}
\fi
 \caption{The RCRS toolset.}
 \label{fig_toolset}
\end{figure}

The RCRS framework comes with a toolset, illustrated in Fig.~\ref{fig_toolset}.
The toolset is publicly available under the MIT license and can be downloaded from \url{http://rcrs.cs.aalto.fi}. 
The toolset is described in tool papers~\cite{RCRS_Toolset_arxiv_2017,RCRS_Toolset_TACAS_2018} (\cite{RCRS_Toolset_arxiv_2017} contains additional material over~\cite{RCRS_Toolset_TACAS_2018}, specifically, a six-page appendix describing a demo of the RCRS toolset).
In summary, the toolset consists of:
\begin{itemize}
\item A full implementation of the RCRS theory in Isabelle~\cite{NipkowPW02}. 
The implementation consists of \fileTotal theory files and a total of \lineTotal lines of Isabelle code.
A detailed description of the implementation can be found in the file \texttt{document.pdf} available in the public distribution of RCRS. 
\item A formal {\em Analyzer}, which is a set of procedures implemented on top of Isabelle and the functional programming language SML. The Analyzer performs
compatibility checking, automatic contract simplification, and other functions.
\item A formalization of Simulink characterizing basic Simulink blocks as
RCRS components and implementing those as a library of RCRS/Isabelle. 
At the time of writing this paper, 48 of Simulink's blocks can be handled.
\item A {\em Translator\/}: a Python program translating Simulink hierarchical block diagrams into RCRS code.
\end{itemize}

We implemented in Isabelle a {\em shallow embedding} \cite{Boulton:1992} of the language 
introduced in \S\ref{sec:syntax}.
The advantage of a shallow embedding is that all datatypes of Isabelle are available for specification of components, and we can use the existing Isabelle mechanism for renaming bound variables in compositions. The disadvantage of this shallow embedding is that we cannot express Algorithm~\ref{fig_atomic_algo} within Isabelle
(hence the ``manual'' proof that we provide for Theorem~\ref{thm:atomic}).
A {\em deep embedding}, in which the syntax of components is defined as a datatype of Isabelle, is possible, and is left as an open future work direction.

We implemented Algorithm~\ref{fig_atomic_algo} in SML, the meta-language of Isabelle. 
The SML program takes as input a component $C$ and returns not only a simplified
atomic component $\symbolic(C)$, but also a proved Isabelle theorem 
of the fact $C\equiv \symbolic(C)$. 
The simplification program, as well as a number of other procedures
to perform compatibility checking, validity checking, etc.,
form what we call the {\em Analyzer} in Fig.~\ref{fig_toolset}.

The {\em Translator} takes as input a Simulink model and produces an 
RCRS/Isabelle theory file containing:
(1) the definition of all atomic and composite components representing
the Simulink diagram; and (2) embedded bottom-up simplification procedures
and the corresponding correctness theorems. 
By running this theory file in Isabelle, we obtain an atomic component 
corresponding to the top-level Simulink model, equivalent to the original
composite component. As a special case, if the Simulink diagram contains
inconsistencies (e.g., division by zero), these are detected by obtaining $\Fail$ 
as the top-level atomic component. The error can be localized by finding
earlier points in the Simulink hierarchy (subsystems) which already
resulted in $\Fail$.

As mentioned earlier, how to obtain a composite component from a graphical
block diagram is an interesting problem. This problem is studied in depth
in~\cite{DragomirPT16}, where several translation strategies are proposed.
These various strategies all yield semantically equivalent components,
but with different trade-offs in terms of size, readability, effectiveness
of the simplification procedures, and so on.
The Translator implements all these translation strategies, allowing the user
to explore these trade-offs.
Further details on the translation problem are provided in~\cite{DragomirPT16,DragomirPreoteasaTripakisFORTE2017}.
A proof that the translation strategies yield semantically equivalent components
is provided in~\cite{PDT:2016}.
This proof, which has been formalized in Isabelle, is a non-trivial result:
the entire formalization of the translation algorithms and the proof is
 \lineTranslate lines of Isabelle code.

We have used the RCRS toolset on several case studies, including a real-life benchmark provided by the Toyota motor company. The benchmark involves a Fuel Control System (FCS) described in \cite{JinDKUB14a,JinDKUB14b}. 
FCS aims at controlling the air mass and injected fuel in a car engine such that their ratio is always optimal. This problem has important implications on lowering pollution and costs by improving the engine performance.
Toyota has made several versions of FCS publicly available as Simulink models at \url{https://cps-vo.org/group/ARCH/benchmarks}.

	We have used the RCRS toolset to process two of the three Simulink models in the FCS benchmark suite (the third model contains blocks that are currently not implemented in the RCRS component library).
A typical model in this set has a 3-layer hierarchy with a total of 104 Simulink block instances (97 basic blocks and 7 subsystems), and 101 connections out of which 8 are feedbacks. 
Each basic Simulink block is modeled in our framework by an atomic STS component
(possibly stateless). These atomic STS components are created once, and form part
of the RCRS implementation, which is reused for different Simulink models.
The particular FCS diagram is translated into RCRS using the Translator,
and simplified within Isabelle using our SML simplification procedure.
After simplification, we obtain an atomic deterministic STS component with no
inputs, 7 outputs, and 14 state variables. 
Its contract (which is 8337 characters long) includes a condition on the state variables, in particular, that a certain state variable must always be non-negative (as its value is fed into a square-root block).
This condition makes
it not immediately obvious that the whole system is valid (i.e., not $\Fail$).
However, we can show after applying the transformation $\sts 2 \qltl$ that
the resulting formula is satisfiable, which implies that the original
model is consistent (i.e., no connections result in incompatibilities, there
are no divisions by zero, etc.).
This illustrates the use of RCRS as a powerful static analysis tool.
More details on the FCS case study are provided in~\cite{DragomirPT16,DragomirPreoteasaTripakisFORTE2017,RCRS_Toolset_arxiv_2017}.

%% file: rwork.tex

\section{Related Work}
\label{sec_related}

Several formal compositional frameworks exist in the literature.
Most closely related to RCRS are the frameworks of
FOCUS~\cite{BroyStolen01},
input-output automata~\cite{LynchTuttle89},
reactive modules~\cite{AlurHenzingerFMSD99},
interface automata~\cite{AlfaroHenzingerFSE01},
and Dill's trace theory~\cite{Dill87}.
RCRS shares with these frameworks many key compositionality principles, such as
the notion of refinement.
At the same time, RCRS differs and complements these frameworks in important ways.
Specifically, FOCUS, IO-automata, and reactive modules, are limited to input-receptive systems, while
RCRS is explicitly designed to handle non-input-receptive specifications. The
benefits of non-input-receptiveness are discussed extensively in~\cite{TripakisLHL11} and will not be repeated here.
Interface automata are a low-level formalism whereas RCRS specifications and
reasoning are symbolic. For instance, in RCRS one can naturally express systems with infinite state-spaces, input-spaces, or output-spaces. (Such systems can
even be handled automatically, provided the corresponding logic they are
expressed in is decidable.) Both interface automata
and Dill's trace theory use a single form of asynchronous parallel composition,
whereas RCRS has three primitive composition operators (serial, parallel, feedback)
with synchronous semantics.

Our work adapts and extends to the reactive system setting many of the ideas
developed previously in a long line of research on correctness and compositionality for sequential programs.
This line of research goes back to the works of Floyd, Hoare, Dijkstra, and
Wirth, on formal program semantics, weakest preconditions, program
development by stepwise refinement, 
and so on~\cite{Floyd67,Hoare69,Dijkstra72,Wirth71}.
It also goes back to game-theoretic semantics of sequential programs
as developed in the original refinement calculus~\cite{backwright:98}, as well as
to contract-based design~\cite{Meyer92}.
Many of the concepts used in our work are in spirit similar to those used
in the above works. For instance, an input-output formula $\phi$ used in
an atomic component in our language can be seen as a {\em contract} between
the environment of the component and the component itself: the environment
must satisfy the contract by providing to the component legal inputs, and
the component must in turn provide legal outputs (for those inputs).
On the other hand, several of the concepts used here come from the world of
reactive systems and as such do not have a direct correspondence in the world 
of sequential programs. For instance, this is the case with feedback composition.

RCRS extends refinement calculus from predicate to property transformers. 
Extensions of refinement calculus to infinite behaviors have also been
proposed in the frameworks of {\em action systems}~\cite{back:wright:1994},
{\em fair action systems}~\cite{back:xu:1998}, and
{\em Event B}~\cite{abrial:2010}. 
These frameworks use predicate (not property) transformers as semantic
foundation; they can handle certain property patterns (e.g., fairness) by
 providing proof rules for these properties, but they 
do not treat livenes and LTL properties in general~\cite{back:xu:1998,butler:divakar:2009,hoang:abrial:2011}.
The {\em Temporal Logic of Actions}~\cite{lamport:1994} can be used to specify 
liveness properties, but does not distinguish between inputs and outputs, and
as such cannot express non-input-receptive components.

Our specifications can be seen as ``rich'', behavioral types~\cite{LeeXiong01,AlfaroHenzingerFSE01}.
Indeed, our work is closely related to programming languages and 
type theory, specifically, {\em refinement types}~\cite{FreemanPfenning1991},
{\em behavioral types}~\cite{Nierstrasz93,LiskovWing94,DharaLeavens96},
and {\em liquid types}~\cite{Rondon2008liquidTypes}.

Behavioral type frameworks have also been proposed in reactive system settings.
In the SimCheck framework~\cite{SimCheck2010}, Simulink blocks
are annotated with constraints on input and output variables, much like
stateless components in RCRS.
RCRS is more general as it also allows to specify stateful components.
RCRS is also a more complete compositional framework, 
with composition operators
and refinement, which are not considered in~\cite{SimCheck2010}.
Other behavioral type theories for reactive systems have been proposed
in~\cite{AlfaroHenzingerEMSOFT01,Chakrabarti02,DoyenHJP08}. Compared
to RCRS, these works are less general. 
In particular, \cite{AlfaroHenzingerEMSOFT01,DoyenHJP08} are limited to
specifications which separate the assumptions on the inputs
from the guarantees on the outputs, and as such cannot capture input-output
relations. \cite{Chakrabarti02} considers a synchronous model which
allows to specify legal values of inputs and outputs at the {\em next} step,
 given the current state. This model does not allow to capture relations
between inputs and outputs within the same step, which RCRS allows.

Our work is related to formal verification frameworks for {\em hybrid systems}~\cite{Hybrid95}. Broadly speaking, these can be classified into frameworks following a
model-checking approach, which typically use automata-based specification languages
and state-space exploration techniques, and those following a theorem-proving
approach, which typically use logic-based specifications.
More closely related to RCRS are the latter, among which,
CircusTime~\cite{Cavalcanti2013}, KeYmaera~\cite{DBLP:conf/cade/FultonMQVP15},
and the PVS-based approach in~\cite{DBLP:conf/iceccs/Abraham-MummSH01}.
CircusTime can handle a larger class of Simulink diagrams than the current
implementation of RCRS. In particular, CircusTime can handle {\em multi-rate}
diagrams, where different parts of the model work at different rates (periods).
On the other hand, CircusTime is based on predicate (not property) transformers,
and as such cannot handle liveness properties.
KeYmaera is a theorem prover based on
{\em differential dynamic logic}~\cite{DBLP:journals/jar/Platzer08}, which is
itself based on {\em dynamic logic}~\cite{Harel2000DL}.
The focus of both KeYmaera and the work in~\cite{DBLP:conf/iceccs/Abraham-MummSH01}
is verification, and not compositionality. For instance, these works do not
distinguish between inputs and outputs and do not investigate considerations 
such as input-receptiveness.
The work of~\cite{ReissigWR17} distinguishes inputs and outputs, but provides
a system model where the output relation is separated from the transition relation,
and where the output relation is assumed to be total, meaning that there exists
an output for every input and current state combination. This does not allow
to specify non-input-receptive stateless components, such as for example
the $\Div$ component from \S\ref{sec:syntax}.

Our component algebra is similar to the algebra of flownomials~\cite{Stefanescu:2000:NA:518304} 
and to the relational model for non-deterministic dataflow~\cite{hildebrandt2004}. 
In~\cite{Courcelle1987}, graphs and graph operations which can be viewed as block diagrams are represented by algebraic expressions and operations, and a complete equational axiomatization of the equivalence 
of the graph expressions is given. This is then applied to flow-charts as investigated in~\cite{SCHMECK1983165}.
The translation of block diagrams in general and Simulink in particular has been
treated in a large number of papers, with various goals, including verification
and code generation (e.g., see~\cite{TripakisSCC05,MeenakshiBR06,ChenDS09,LublinermanTripakisPOPL09,SfyrlaTSBS10,Bostrom11,YangV12,ZhouK12,ZouZWFQ13,MinopoliF16}).
Although we share several of the ideas of the above works, our main goal here
is not to formalize the language of block diagrams, neither their translation
to other formalisms, but to provide a complete compositional framework for 
reasoning about reactive systems.

RCRS is naturally related to {\em compositional verification} frameworks, such as~\cite{Grumberg94,AbadiLamport95,McMillan97,ShankarLazyCompo97,HenzingerQadeerRajamaniCAV98,1998compositionality,deRoever2012}.
In particular, compositional verification frameworks often make use of a refinement relation such as trace inclusion or simulation~\cite{ShankarLazyCompo97,HenzingerQadeerRajamaniCAV98}.
However, the focus of these frameworks is different than that of RCRS.
In compositional verification, the focus is to ``break down'' a large (and
usually computationally expensive) verification task into smaller (and
hopefully easier to calculate) subtasks. For this purpose, compositional
verification frameworks employ several kinds of {\em decomposition rules}.
An example of such a rule is the so-called {\em precongruence} rule
(i.e., preservation of refinement by composition):
if $P_1$ refines $Q_1$, and $P_2$ refines $Q_2$, then the composition 
$P_1\| P_2$ refines $Q_1\| Q_2$. This, together with preservation
of properties by refinement, allows us to conclude that $P_1\| P_2$ satisfies
some property $\phi$, provided we can prove that $Q_1\| Q_2$ satisfies $\phi$.
The latter might be a simpler verification task, if $Q_1$ and $Q_2$ are
smaller than $P_1$ and $P_2$.
The essence of compositional verification is in finding such {\em abstract}
versions $Q_1$ and $Q_2$ of the concrete processes in question, $P_1$ and $P_2$,
and employing decomposition rules like the one above in the hope of making
verification simpler. 
RCRS can also be used for compositional verification: indeed, RCRS provides
both the precongruence rule, and preservation of properties by refinement.
Note that, in traditional settings, precongruence is not always powerful enough, 
and for this reason most compositional
verification frameworks employ more complex decomposition rules (e.g.,
see~\cite{NamjoshiTrefler10}).
In settings which allow non-input-receptive components, such as ours,
there are indications that the precongruence rule is sufficient for compositional
verification purposes~\cite{STHACMTECS2017}, although more work is required
to establish this in the specific context of RCRS. Such work is beyond the
scope of the current paper. We also note that, beyond compositional
verification with precongruence, RCRS provides a behavioral type theory which
allows to state system properties such as compatibility, which is typically
not available in compositional verification frameworks.

Refinement can be seen as the inverse of {\em abstraction}, and as such our
framework is related to general frameworks such as {\em abstract interpretation}~\cite{Cousot77}. Several abstractions have been proposed in reactive system settings,
including {\em relational} abstractions for hybrid systems, which are related
to Simulink~\cite{Sankaranarayanan2011}. The focus of these works is verification,
and abstraction is used as a mechanism to remove details from the model that make
verification harder. In RCRS, the simplification procedure that we employ can be
seen as an abstraction process, as it eliminates internal variable information.
However, RCRS simplification is an {\em exact} abstraction, in the sense that
it does not lose any information: the final system is equivalent to the original
one, and not an over- or under-approximation, as is usually the case with
typical abstractions for verification purposes.

%% file: conclusion.tex

\section{Conclusion}
\label{sec:conclusion}

We presented RCRS, a compositional framework for modeling and reasoning about
reactive systems. This paper focuses on the theory and methodology of RCRS,
its formal semantics, and techniques for symbolic and computer-aided reasoning. 
\delete{
We also briefly presented the RCRS toolset. For more information
about the toolset we refer the reader to the relevant papers~\cite{DragomirPT16,PDT:2016,DragomirPreoteasaTripakisFORTE2017,RCRS_Toolset_arxiv_2017,RCRS_Toolset_TACAS_2018}, as well as the toolset's web site \url{http://rcrs.cs.aalto.fi} for
up-to-date information.
}

RCRS is an ongoing project, and a number of problems remain open. Future work
directions include:
\begin{itemize}
\item An extension of the framework to systems with algebraic loops, which
necessitates handling instantaneous feedback. Here, the preliminary ideas
of~\cite{PreoteasaTripakisLICS2016} can be helpful in defining the semantics
of instantaneous feedback. However, \cite{PreoteasaTripakisLICS2016} does not
provide solutions on how to obtain 
symbolic closed-form expression for the feedback of general components.
\item Extension of the results of \S\ref{sec_algo_feedback} to general
components, possibly non-deterministic or non-decomposable.
\item An extension of the framework to {\em acausal} systems, i.e., systems
without a clear distinction of inputs and outputs~\cite{Fritzson2014}.
\item The development of better symbolic reasoning techniques, such as
simplification of logical formulas, decision procedures, etc.
\end{itemize}